\documentclass[fleqn,usenatbib]{mnras}
\usepackage{float}

\usepackage{newtxtext,newtxmath}
\usepackage{graphicx}	% Including figure files
\usepackage{amsmath}	% Adv\usepackage{longtable}anced maths commands
\usepackage{amssymb}	% Extra maths symbols
\usepackage{float}
\usepackage{subfigure}
\usepackage{ulem}
\usepackage{array}
\usepackage{booktabs}
\usepackage{supertabular}
\usepackage{multicol}

\newcommand{\sigfrac}{$\sigma_{\rm frac}$}
\title[The first 62 AGN  in MaNGA: kinematics]{The first 62 AGN observed with SDSS-IV MaNGA -- III: stellar and gas kinematics }

\author[Ilha et al.]{Gabriele S. Ilha,$^{1,2}$\thanks{E-mail: gabrieleilha1994@gmail.com} Rogemar A. Riffel,$^{1,2}$ Jaderson S. Schimoia,$^{1,2,3}$ 
\newauthor  Thaisa Storchi-Bergmann,$^{3,2}$ Sandro B. Rembold,$^{1,2}$ Rog\'erio Riffel,$^{3,2}$   \newauthor Dominika Wylezalek,$^{4}$ Yong Shi,$^{5}$ Luiz N. da Costa,$^{2,6}$ Alice D. Machado,$^{1,2}$  
\newauthor David R. Law,$^{7,8}$   Dmitry Bizyaev,$^{9,10}$  Nicolas D. Mallmann,$^{3,2}$ Janaina Nascimento,$^{3,2}$  
\newauthor Marcio A. G. Maia,$^{2,6}$  Rafael Cirolini$^{1,2}$  \\
$^{1}$Departamento de F\'isica, CCNE, Universidade Federal de Santa Maria, 97105-900, Santa Maria, RS, Brazil\\
$^{2}$Laborat\'orio Interinstitucional de e-Astronomia - LIneA, Rua Gal. Jos\'e Cristino 77, Rio de Janeiro, RJ - 20921-400, Brazil\\
$^{3}$Departamento de F\'isica, IF, Universidade Federal do Rio Grande do Sul, CP 15051, 91501-970, Porto Alegre, RS, Brazil\\
$^{4}$European Southern Observatory, Karl-Schwarzschildstr 2, 85748 Garching bei M\"unchen, Germany\\
$^{5}$Department of Astronomy, Nanjing University, Nanjing 210093, China	\\
$^{6}$Observat\'orio Nacional - MCT, Rua General Jos\'e Cristino 77, Rio de Janeiro, RJ - 20921-400, Brazil\\
$^{7}$Space Telescope Science Institute, 3700 San Martin Drive, Baltimore, MD 21218, USA	\\
$^{8}$Dunlap Institute for Astronomy and Astrophysics, University of Toronto, 50 St. George Street, Toronto, Ontario M5S 3H4, Canada\\
$^{9}$Apache Point Observatory, PO Box 59, Sunspot, NM 88349, USA \\ 
$^{10}$Sternberg Astronomical Institute, Moscow State University, 119992 Moscow, Russia
}

\date{Accepted XXX. Received YYY; in original form ZZZ}

\pubyear{2016}

\begin{document}
\label{firstpage}
\pagerange{\pageref{firstpage}--\pageref{lastpage}}
\maketitle

% * <jaderfisico@gmail.com> 2018-01-23T15:16:11.515Z:
% 
% 
% ^.
\begin{abstract}
%Theoretical studies and numerical simulations suggest that Active Galactic Nuclei (AGN) play an important role on the evolution of their host galaxies, then AGN feedback is important to understand this evolution. In this work, w
We investigate the effects of Active Galactic Nuclei (AGN) on the gas kinematics of their host galaxies, using MaNGA data for a sample of 62 AGN hosts and 109 control galaxies (inactive galaxies). We compare orientation of the line of nodes (kinematic Position Angle -- PA) measured from the gas and stellar velocity fields for the two samples. We found that AGN hosts and control galaxies display similar kinematic PA offsets between gas and stars. However, we note that AGN have larger fractional velocity dispersion $\sigma$ differences between gas and stars [$\sigma_{\rm frac}=(\sigma_{\rm gas}-\sigma_{\rm stars})/\sigma_{\rm stars}$] when compared to their controls, as obtained from the velocity dispersion values of the central (nuclear) pixel (2\farcs5 diameter). The AGN have a median value of $\sigma_{\rm frac}$ of $<\sigma_{\rm frac}>_{\rm AGN}=0.04$, while the the median value for the control galaxies is $<\sigma_{\rm frac}>_{\rm CTR}=-0.23$. 75\,\% of the AGN show $\sigma_{\rm frac}>-0.13$, while 75\,\% of the normal galaxies show $\sigma_{\rm frac}<-0.04$, thus  we suggest that the parameter  \sigfrac\  can be used as an indicative of AGN activity. We find a correlation between the [O\,{\sc iii}]$\lambda$5007 luminosity and $\sigma_{\rm frac}$ for our sample.  Our main conclusion is that the AGN already observed with MaNGA are not powerful enough to produce important outflows at galactic scales, but at 1-2 kpc scales, AGN feedback signatures are always present on their host galaxies.
\end{abstract}


%%%%%%%%%%%%%%%%%%%%%%%%%%%%%%%%%%%%%%%%%%%%%%%%%%

%%%%%%%%%%%%%%%%% BODY OF PAPER %%%%%%%%%%%%%%%%%%

\section{Introduction}

Theoretical studies and numerical simulations suggest that Active Galactic Nuclei (AGN) play an important role in the evolution of their host galaxies \citep[e.g.][]{hopkins05}. Currently, it is widely accepted that galaxies with spherical component (bulge of spiral galaxies and elliptical galaxies) host a central supermassive black hole 
 \citep[SMBH,][]{ferrarese00,gebhardt00,tremaine02,scannapieco05} and cosmological simulations that do not include  feedback effects from the SMBH result in galaxy stellar masses much higher than observed  \citep{diMatteo05,springel05,bower06}. Massive outflows originated in the accretion flow are claimed to regulate and couple the growth of the  galactic bulge and SMBH \citep{hopkins05} and to explain the relation between the mass of the SMBH and stellar velocity dispersion of the bulge -- the $M-\sigma$ relation \citep[e.g.][]{ferrarese00,gebhardt00}.%,  as they prevent  the accretion of extragalactic gas in active phases of the galaxy \citep{nemmen07}.

 According to the Unified Model for AGN \citep[e.g.][]{antonucci93,urry95}, the Narrow Line Region (NLR) is expected to present a bi-conical shape, within which gas outflows due to winds from the accretion disk are expected to be observed. However, Hubble Space Telescope (HST) narrow-band [O\,{\sc iii}]$\lambda$5007 images of a sample of 60 nearby Seyfert galaxies show that the bi-conical shape of the NLR is not as common as expected \citep{schmitt03} and gas outflows are seen only in 33\% of Seyfert galaxies, as revealed by long-slit spectroscopy of 48 nearby AGN \citep{fischer13}. Nevertheless, long-slit observations are restricted to only one position angle. A better mapping of the outflows and their geometries can be obtained via integral field spectroscopy (IFS), as shown in recent studies both in the optical and near-infrared  \citep[e.g.][]{lena15,allan14,zakamska16, n5929let,barbosa14}.%%integral field spectroscopy (IFS) of nearby active galaxies, suggesting that the spatial coverage of IFS is essential to properly map the gas kinematics and look for outflows from AGN. 
 
 The comparison between the gas and stellar kinematics on kiloparsec scales allows the study of the possible impact of AGN outflows on its host galaxy. So far, most studies aimed to investigate gas outflows from AGN have been performed for small samples or individual galaxies. In this work we  use the observations from the Mapping Nearby Galaxies at the Apache Point Observatory (MaNGA) survey \citep{bundy15} to compare the gas and stellar kinematics of a sample composed by 62 AGN observed in the MPL-5 (MaNGA Product Launch V) \citep[Data Release 14,][]{dr14} with those  of a control sample of inactive galaxies, matched with the AGN sample by properties of the host galaxies. If an AGN  sample presents strong outflows, the large-scale gas velocity fields are expected to be disturbed when compared to the stellar velocity fields, while for inactive galaxies, the stellar and gas velocity fields are expected to be similar. Another way that AGN can affect the gas dynamics is by increasing the gas velocity dispersion due to the shocks of the nuclear outflow with the ambient gas. 
 
 The AGN and control samples used in this paper are described in \citet{rembold17} (hereafter Paper I), which presents also the study of the nuclear stellar populations. %The control sample is composed by two galaxies for each AGN, selected to match the properties of the AGN host:  stellar mass, redshift, visual morphology and inclination. 
This is the third paper of a series aimed to compare properties of AGN hosts and their control galaxies. Besides Paper I, the spatially resolved stellar populations is investigated in \citet{nicolas18} (Paper II). In addition, the gas excitation and distribution will presented by Nascimento et al. (in preparation -- Paper IV).

% ***Here I believe we should write a little bit more about the motivation of the paper, or how the outflows are identified and seen (outflow signatures) in the host galaxy*** 
 
 This paper is organized as follows: Section~\ref{data} presents the samples of active and inactive galaxies and the data analysis methods, while Section~\ref{results} presents the results, which are discussed in Section~\ref{discussion}. Finally, the conclusions of this work are presented in Section~\ref{conclusions}.

\begin{figure*}
\begin{center}
1-339163
\end{center}
\centering
\includegraphics[width=0.99\textwidth]{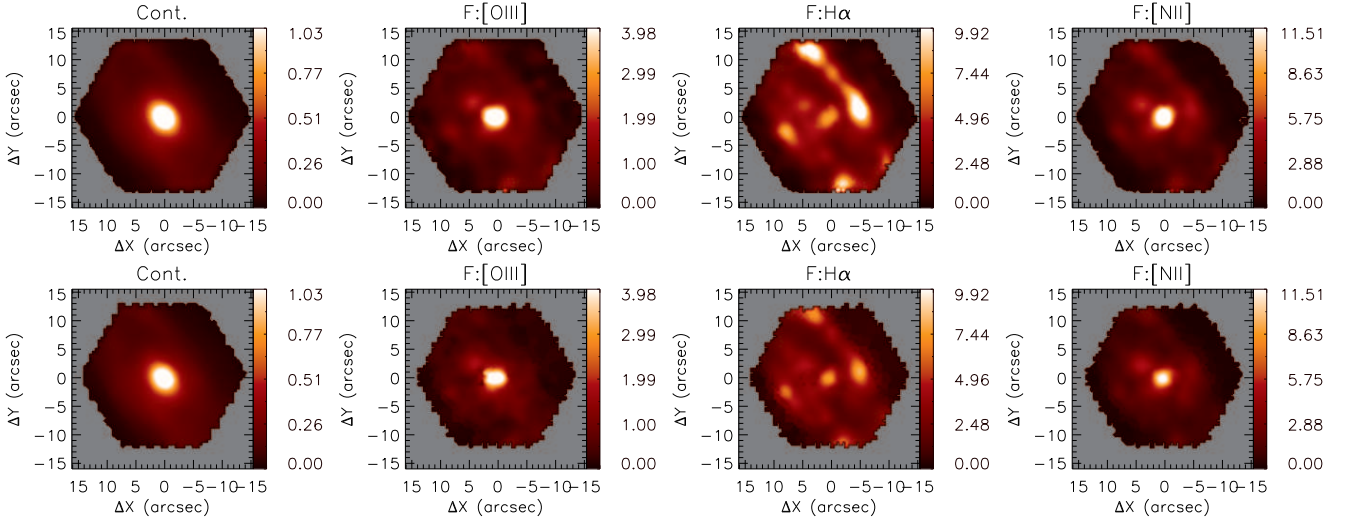}
\caption{Emission lines fluxes for the galaxy with {\it mangaid} 1-339163. Our measurements are shown at the top row and the MaNGA-DAP measurements at the bottom row. In all panels, the North points up and East to the left and the $x$ and $y$ labels show the distance relative to the peak of the continuum emission. The first column shows  a map of the continuum emission obtained by collapsing the whole spectral range, the following columns exhibit the spatial distribution of the emission line fluxes for [O\,{\sc iii}]5007\,\AA, H$\alpha$ and [N\,{\sc ii}]6583\,\AA, respectively. The color bars show the fluxes in unit of 10$^{-17}$ erg s$^{-1}$ cm$^{-2}$ spx$^{-1}$.} 
\label{Fig.1}
\end{figure*}

\begin{figure*}
\begin{center}
1-339163
\end{center}
\centering
\includegraphics[width=0.99\textwidth]{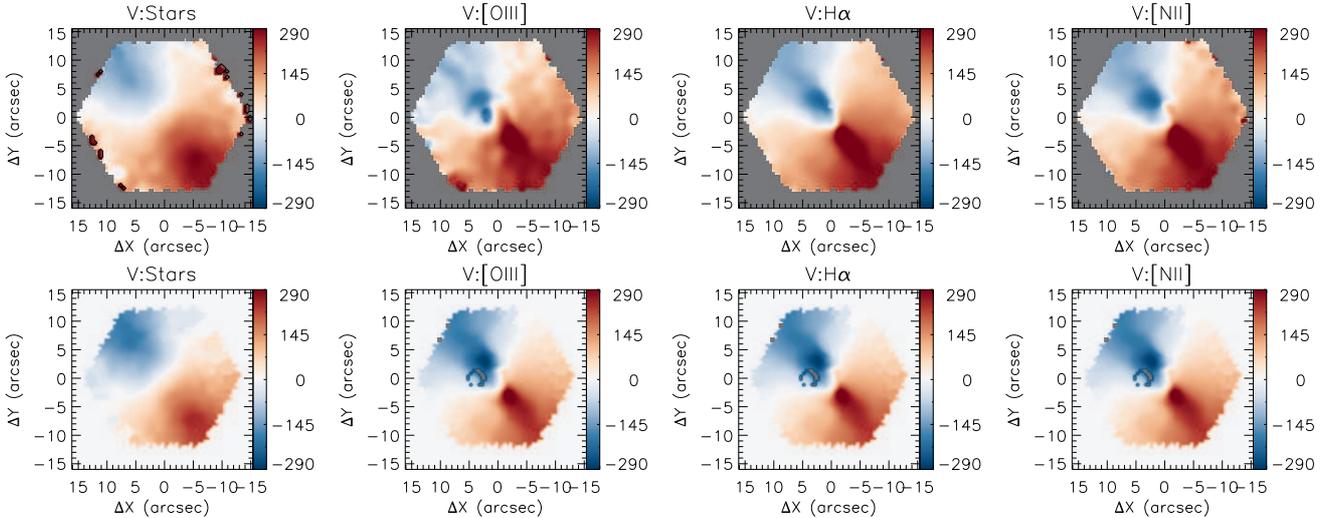}
\caption{Velocity fields for the galaxy {\it mangaid} 1-339163. Our measurements are shown at the top row and the DAP measurements at the bottom row. In all panels, the North points up and East to the left and the $x$ and $y$ labels show the distance relative to the peak of the continuum emission. The systemic velocity has been subtracted from each panel. The first column shows the stellar velocity field and the following columns exhibit the velocity fields for [O\,{\sc iii}], H$\alpha$ and [N\,{\sc ii}], respectively. The velocity maps are in unit of km s$^{-1}$ relative to the systemic velocity of the galaxy.} 
\label{Fig.2}
\end{figure*}

\begin{figure*}
\begin{center}
1-339163
\end{center}
\centering
\includegraphics[width=0.99\textwidth]{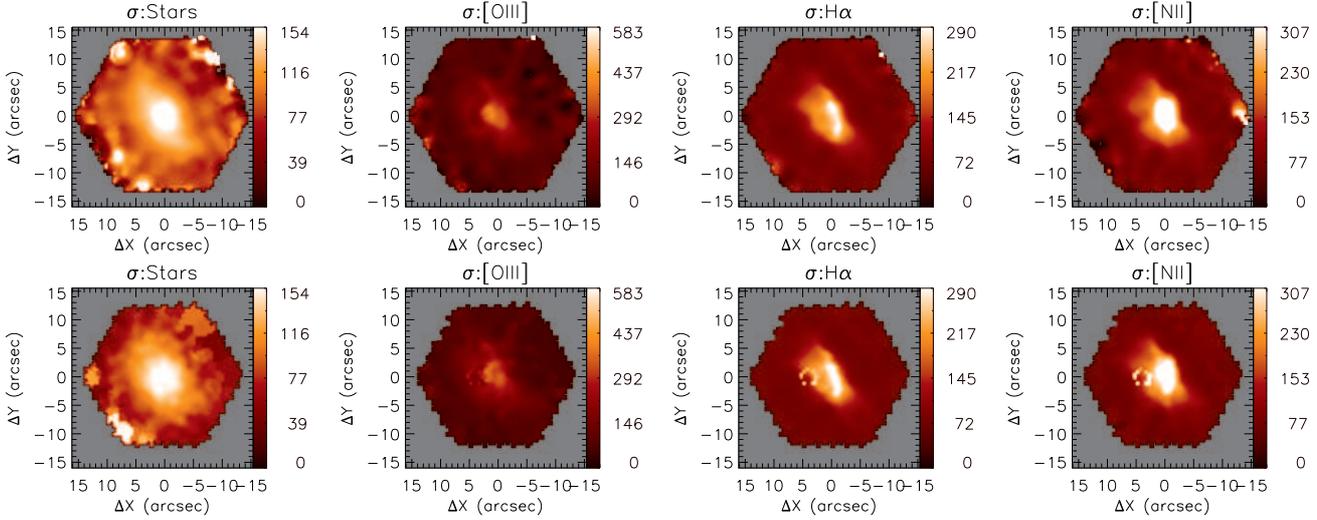}
\caption{Velocity dispersion maps for the galaxy with {\it mangaid} 1-339163. Our measurements are shown at the top row and the DAP measurements at the bottom row. In all panels, the North points up and East to the left and the $x$ and $y$ labels show the distance relative to the peak of the continuum emission. The first column shows the stellar velocity dispersion distribution and the following columns exhibit the gas velocity dispersion distributions for [O\,{\sc iii}], H$\alpha$ and [N\,{\sc ii}], respectively. The color bars show the velocity dispersion corrected by instrumental broadening in units of  km s$^{-1}$.} 
\label{Fig.3}
\end{figure*}

\begin{figure*}
\centering
\begin{center}
1-95092
\end{center}
\includegraphics[width=\textwidth]{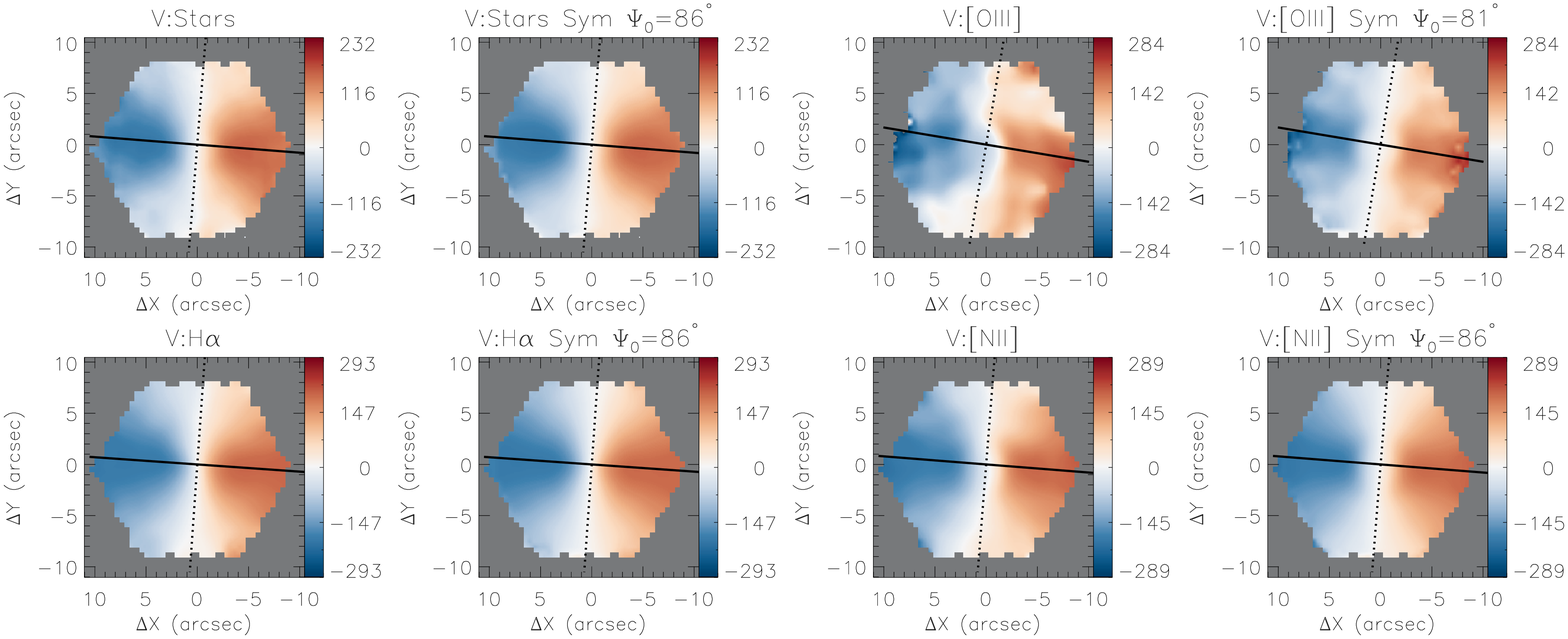}
\begin{center}
1-351790
\end{center}
\includegraphics[width=\textwidth]{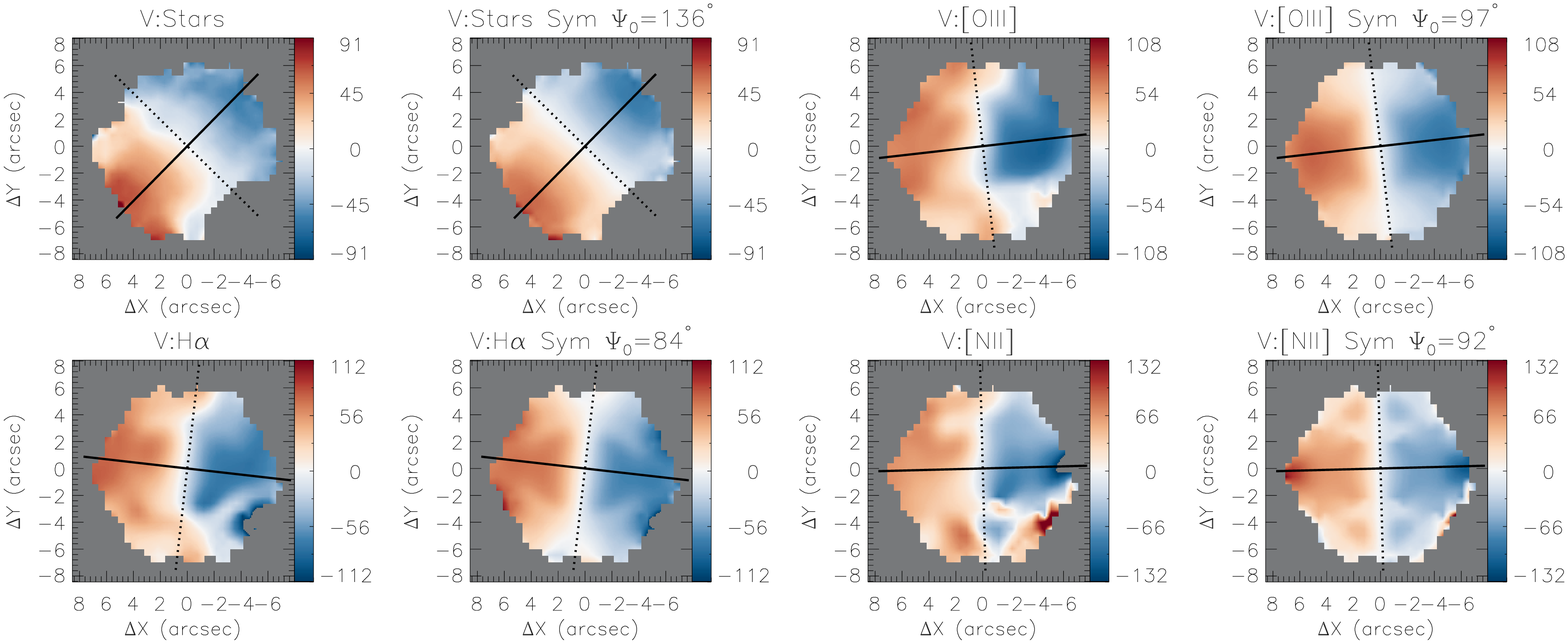}
\caption{
In the first two rows we show the derived velocity fields for the AGN {\it mangaid} 1-95092. The first row shows, from left to right,  the stellar velocity field (V:Stars), the symmetrized stellar velocity field (V:Stars Sym), the gas velocity field for [O\,{\sc iii}] (V:[{O}{\sc iii}]), and the corresponding symmetrized velocity field  (V:[{O}{\sc iii}] Sym).  The second row shows, from left to right, the H$\alpha$ velocity field (V:H$\alpha$), its symmetrized velocity field (V:H$\alpha$ Sym), the velocity and symmetrized velocity fields for  [N\,{\sc ii}]   (V:[{N}{\sc iii}]) and   (V:[{N}{\sc ii}] Sym), respectively. In the bottom two rows we show the same velocity maps but for the AGN {\it mangaid} 1-351790.
In all velocity maps the solid black line shows the position angle of kinematic major axis, the value of the $\Psi_0$ is shown in the top right-corner of the symmetrized velocity maps. The continuous (dotted) line shows the orientation of the kinematic major (minor) axis of the galaxy. The color bars show the velocities in units of km s$^{-1}$.
}
\label{Fig.4}
\end{figure*}

\begin{figure*}
\centering
\begin{center}
12-129446
\end{center}
\includegraphics[width=\textwidth]{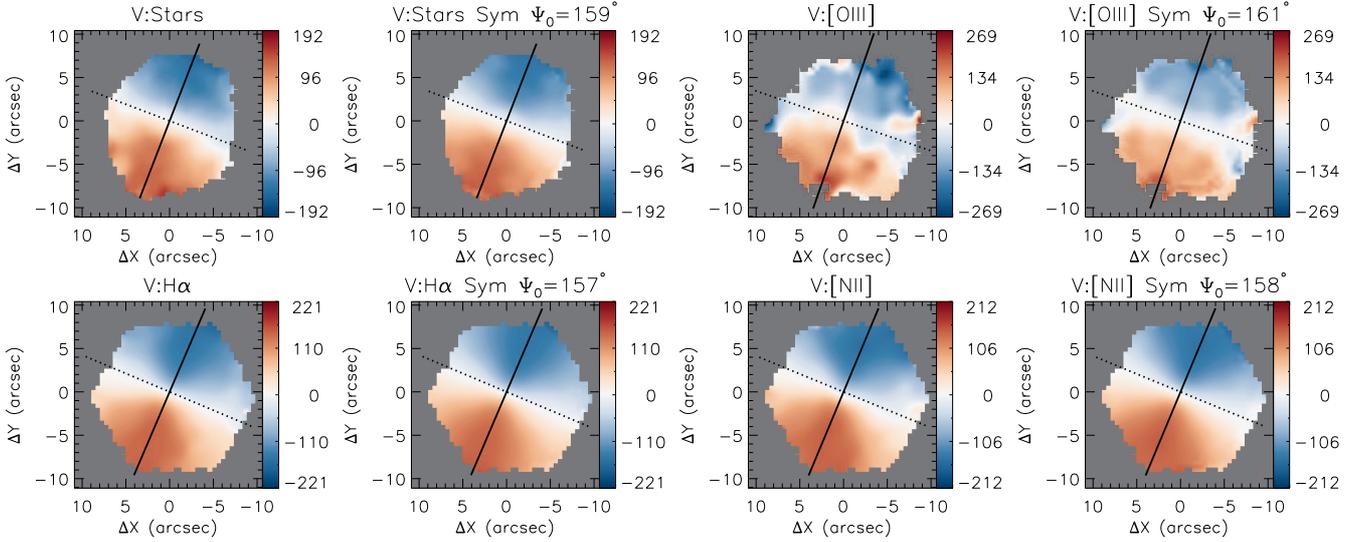}
\begin{center}
1-178838
\end{center}
\includegraphics[width=\textwidth]{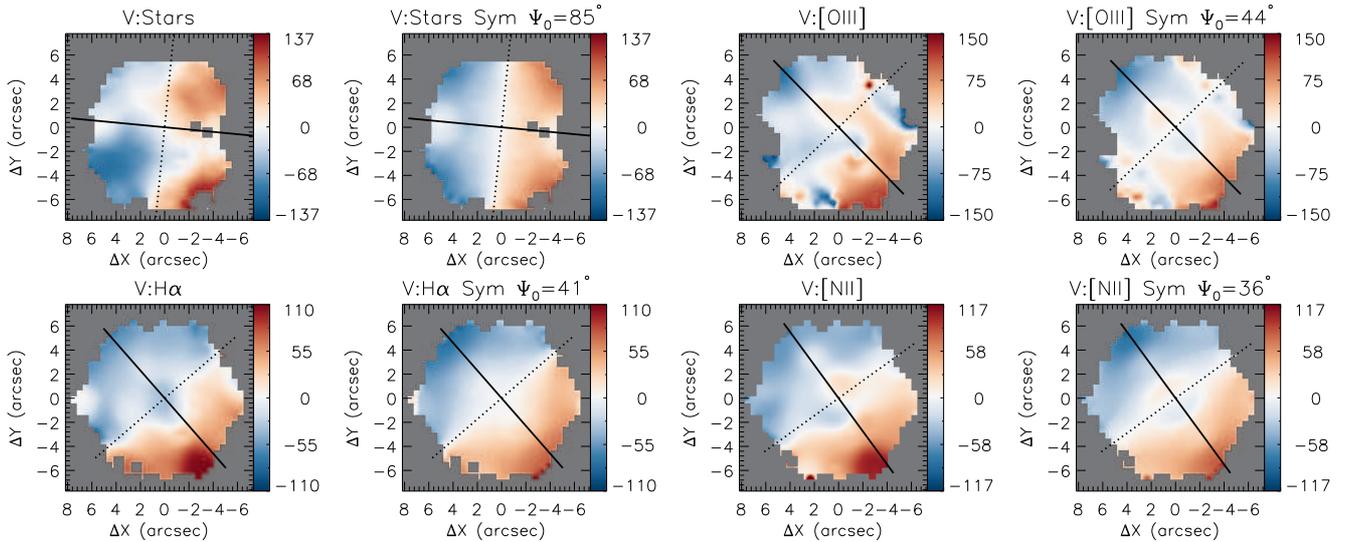}
\caption{Same as Fig.\,\ref{Fig.4} for the control galaxies {\it mangaid} 12-129446 (first and second row) and {\it mangaid} 1-178838 (third and fourth row). The velocity fields are in unit of km s$^{-1}$.} 
\label{Fig.5}
\end{figure*}

\begin{figure}
  \centering
  \includegraphics[width=0.49\textwidth]{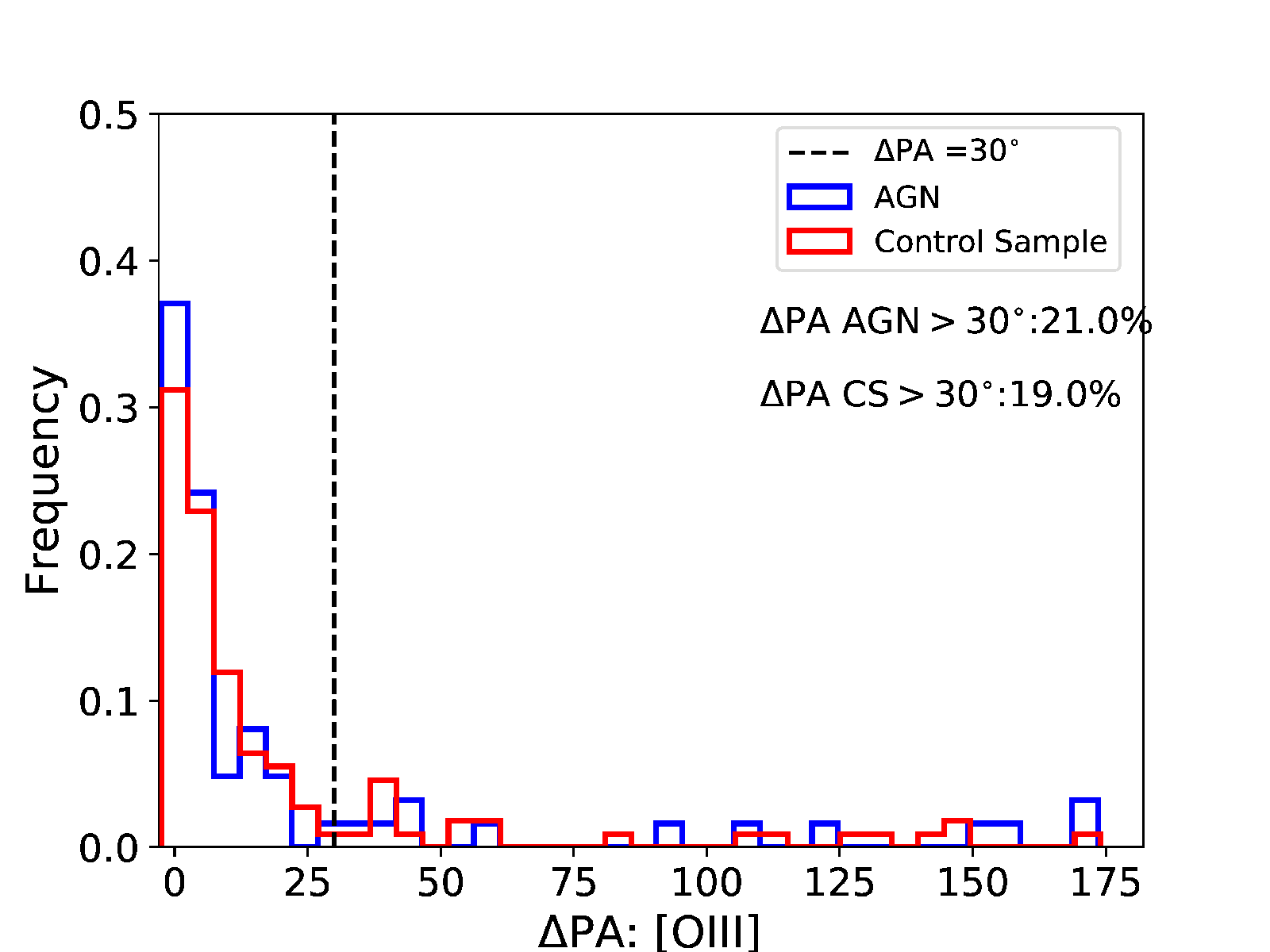}
 \includegraphics[width=0.49\textwidth]{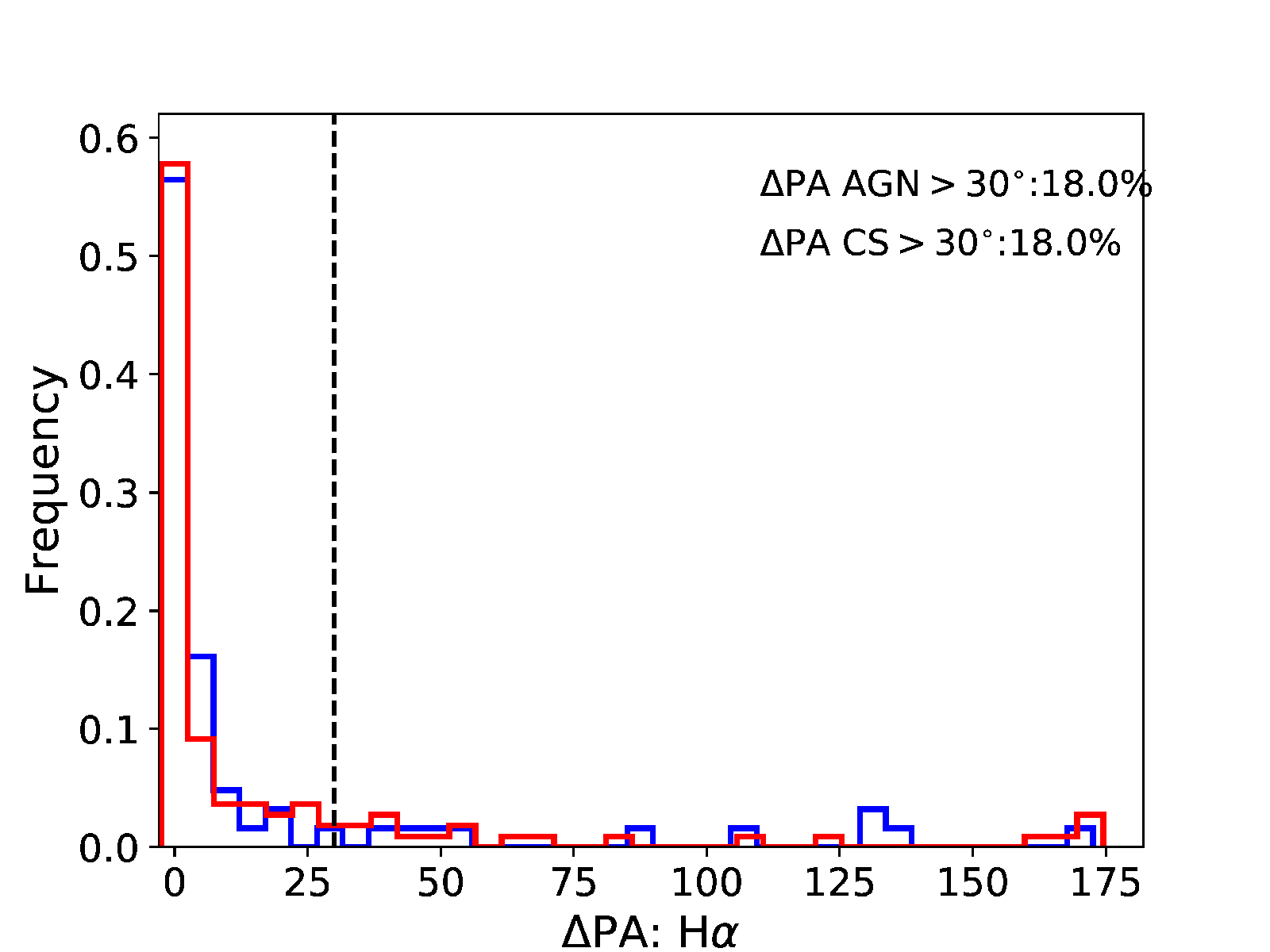}
  \includegraphics[width=0.49\textwidth]{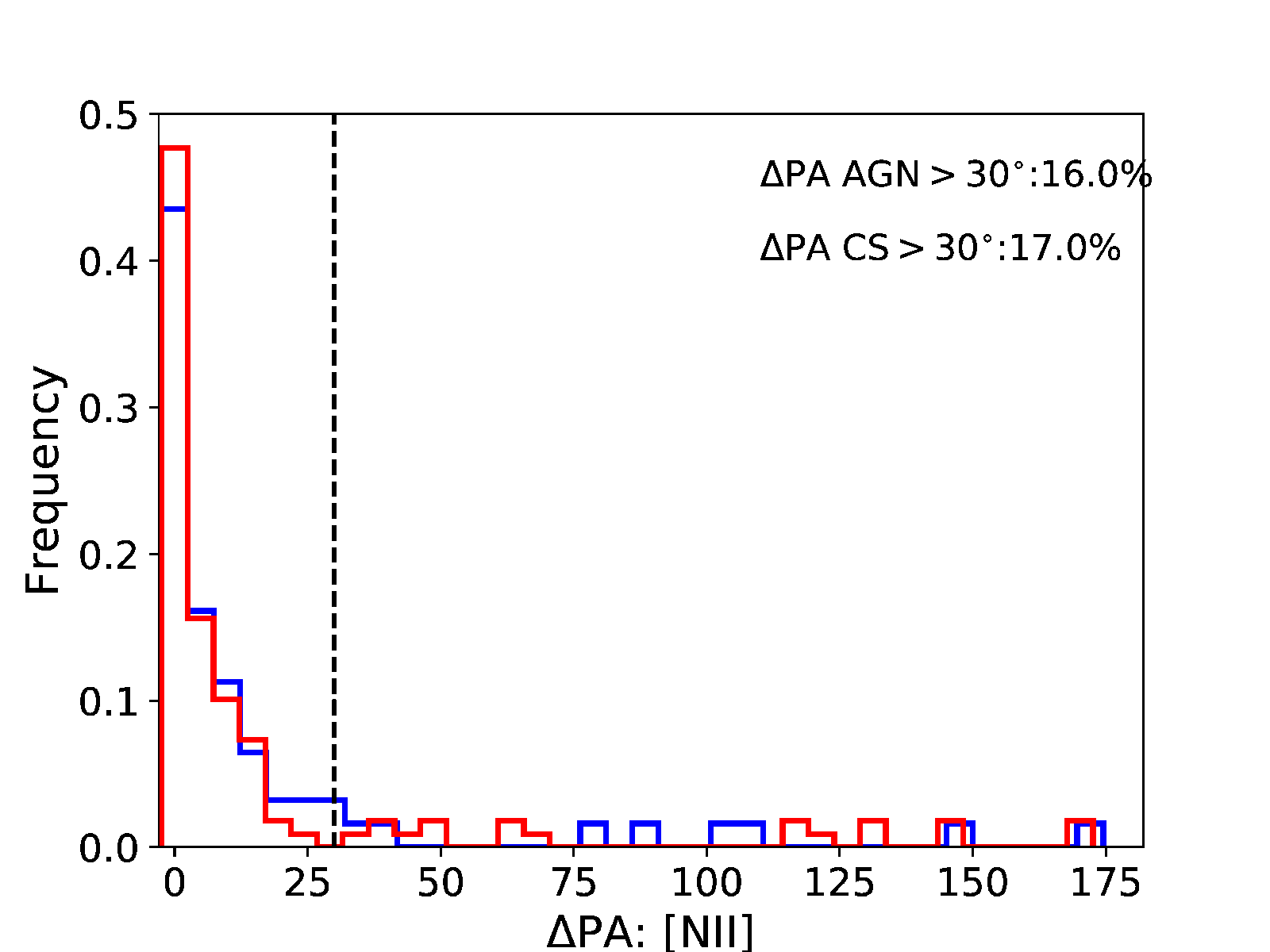}
  \caption{Histograms comparing $\Delta$PA distributions of AGN and control galaxies for [O\,{\sc iii}]$\lambda$5007 (top panel), H$\alpha$ (middle panel) and [N\,{\sc ii}]$\lambda$6583 (bottom line). AGN are shown in blue and controls in red. The vertical dashed lines show $\Delta$PA=30$^\circ$. }
  \label{Fig.6}
  \end{figure}

 \begin{figure}
  \centering
\includegraphics[width=0.49\textwidth]{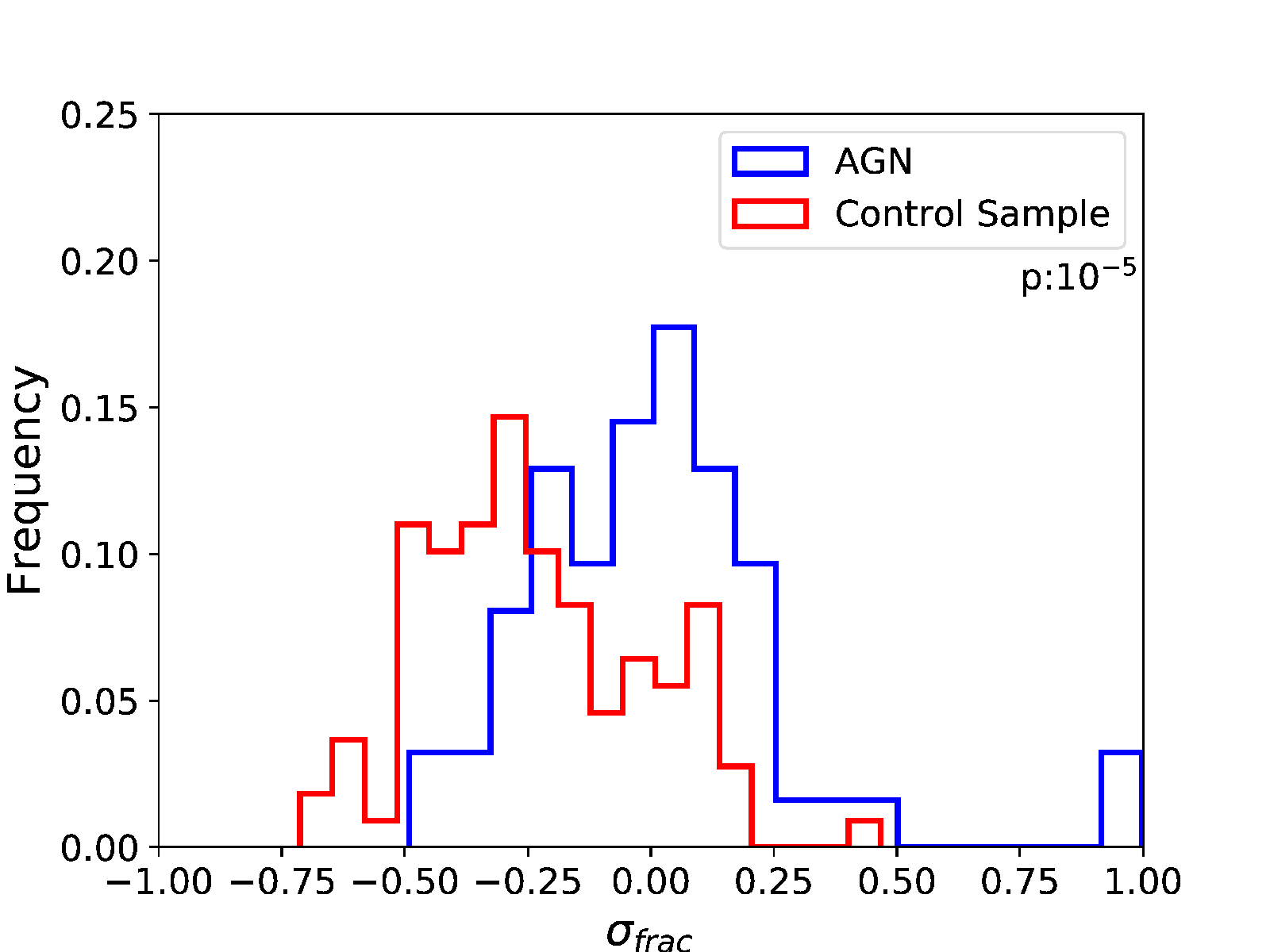}
  \caption{Histograms comparing $\sigma_{frac}$ distributions of AGN and control galaxies for [O\,{\sc iii}]$\lambda$5007. AGN are shown in blue and controls in red. The result of Anderson-Darling statistical test returns a p-value of $10^{-5}$, confirming that  AGN and inactive galaxies follow distinct distributions in $\sigma_{\rm frac}$.}
  \label{Fig.7}
\end{figure}
  
\begin{figure}
  \centering
\includegraphics[width=0.49\textwidth]{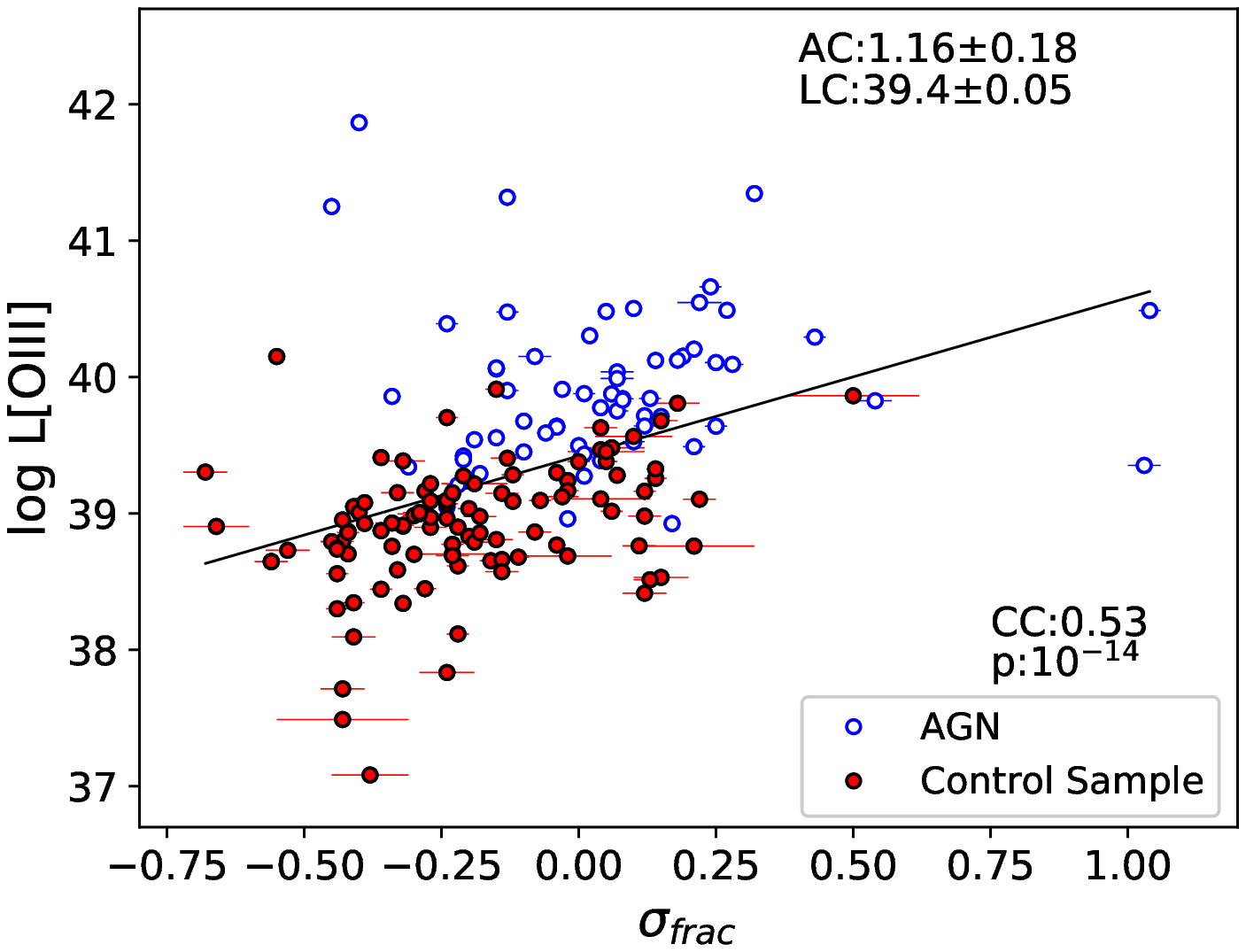}
  \caption{Plot of logarithm of [O\,{\sc iii}]5007\,\AA\ luminosity versus $\sigma_{\rm frac}$ for AGN (blue open circles) and inactive galaxies (red closed circles). The Spearman test confirms that these properties are correlated resulting in a correlation coefficient of 0.53 and p-value of 10$^{-14}$. The black line is the result of linear fit of data, with linear coefficient of $LC=39.4\pm0.05$ and angular coefficient of $AC=1.16\pm0.18$.}
  \label{Fig.8}
\end{figure}

\section{The Data and analysis } \label{data}

\subsection{Sample and MaNGA data}

%***Here I believe it is better to explain the MaNGA survey and data and  then speak about the sample. 

We  use the  datacubes obtained within the MaNGA survey of the sample of AGN and matched control sample defined in Paper I.  The MaNGA survey is part of the fourth-generation Sloan Digital Sky Survey (SDSS-IV) and is aimed to observe $\sim$ 10,000 nearby galaxies using optical Integral Field Spectroscopy (IFS) covering the spectral range  3600--10000~\AA\ and spectral resolving power $R\sim2000$ at a spatial resolution of 1--2 kpc. The MaNGA sample of galaxies was designed to cover at least  $1.5\,R_e$ ($R_e$ -- effective radius). Here the  effective radius is defined as the radius that contains the half luminosity of galaxy measured at the i-band as described in \cite{bundy15}. The MaNGA survey science goals are presented in \citet{bundy15},  the design and performance of the Integral Field Units are discussed in \citet{drory15} and the MaNGA sample is presented in \citet{wake17}.   \citet{yan16b} present the survey design, execution, and data quality, the observing strategy is presented in \citet{law15} and  the data reduction and calibrations are discussed in \citet{law16} and \citet{yan16}. 

Our sample is composed by the first 62 AGN observed with MaNGA -- selected from MaNGA MPL-5 \citep[Data Release 14,][]{dr14}. For each AGN, two control inactive galaxies, matched to the AGN hosts in absolute magnitude, galaxy mass, redshift, morphological type and inclination, were selected. The AGN selection realized by  \citet{rembold17}  is based on single-fiber SDSS-III observations. A detailed description and characterization of the AGN and control samples is presented in Paper I and the properties of AGN and control galaxies are show in Table \ref{tableagns}  and \ref{tablecontrol}, respectively. \citet{wylezalek18} found 173 galaxies that would not have been selected as AGN candidates based on single-fiber spectral measurements, but MaNGA allowed AGN selection based on the fully spatially resolved optical diagnostics and in the future papers similar work will be done for ``nuclear'' AGN and ``off-nuclear'' AGN. Thus, in this work we focus on the ``nuclear'' AGN. As mentioned in \citet{rembold17}, our AGN sample includes  34  (55  per  cent) spiral  and  18  (29  per  cent)  elliptical  galaxies. The  remaining  10  objects  (16  per  cent)  comprise  6  E/S  galaxies,  1 merger  and  3  unclassified  objects.

\subsection{Spatial filtering and noise removal}

In order to remove noise from the observed datacubes, without loss of angular resolution, we performed a spatial filtering of the datacubes using a Butterworth bandpass filter \citep{gonzalez02}. This filter is performed in the frequency domain. We used a low-bandpass filter to remove high spatial frequency components from the cubes, which are usually due to spurious features (e.g. bad pixels or cosmic rays). This procedure allow us to improve the fit the emission and absorption line spectra, as compared with the original datacubes.

 To perform the spatial filtering, we used the Interactive Data Language (IDL) routine $bandpass_-filter.pro$, which allows the choice of the cut-off frequency ($\nu$) and the order of the filter $n$. A low value of $n$ (e.g. 1) is close to a Gaussian filter, while a high value (e.g. 10) corresponds to an Ideal filter.  We used $n=5$ and $\nu=0.25$~Ny, chosen by comparing the filtered cubes with the original ones. For lower values of  $\nu$, besides the removal of spatial noise, the filter excludes also emission from the nucleus of the galaxy. % \citet{menezes15} present a detailed description of the spatial filtering procedure, although the details of the implementation of the spatial filtering used here are distinct than that used by these authors.

\subsection{Spectral fitting}

In order to measure the emission-line fluxes and the stellar and gas kinematics from the MaNGA datacubes, we used the Gas AND Absorption Line Fitting ({\sc gandalf}) code \citep{sarzi06,oh11}.  In brief, the {\sc gandalf} code fits the emission and absorption lines simultaneously, allowing  the separation of  the relative contribution of the stellar continuum and of nebular emission in the spectra of the galaxies. To subtract the underlying stellar contribution on the spectra of the galaxy and measure the stellar kinematics, {\sc gandalf} uses the Penalized Pixel-Fitting ({\sc ppxf}) routine \citep{cappellari04,cappellari17}. The continuum spectra of the galaxy is fitted by using a library of template spectra under the assumption that the line-of-sight velocity distribution (LOSVD) of the stars is well reproduced by a Gauss-Hermite series. 

As template spectra, we used 30 selected Evolutionary Population Synthesis models from
\citet{bc03}, covering ages ranging from 5~Myr to 12~Gyr and three metallicities ($0.004\,Z_{\odot},0.02\,Z_{\odot},0.05\,Z_{\odot}$). During the fit of the spectra, we allowed the use of an order 3 multiplicative Legendre polynomial to correct the shape of the continuum and only the first two Gauss-Hermite moments (velocity and velocity dispersion) were included to represent the LOSVD. We have tested the inclusion of higher order moments, but achieved the best results in the fitting process by considering only the first and second moments.

The emission-line profiles were fitted by Gaussian curves, by keeping tied the centroid velocity and width of the [N\,{\sc ii}]$\lambda\lambda6548,6583$ and [S\,{\sc ii}]$\lambda\lambda6716,6731$  emission lines, fitting each doublet separately. In addition, the following line flux-ratio was kept fixed to their theoretical value: [N\,{\sc ii}]$\lambda6583$/[N\,{\sc ii}]$\lambda6548=2.94$ \citep{osterbrock06}. {\sc gandalf} gives as output measurements for the centroid velocity and velocity dispersion ($\sigma$) of the stars, and the flux, centroid velocity and $\sigma$ of the emission lines  for each spaxel, used to construct two-dimensional maps.

\subsection{Measurements of the Kinematic Position Angles}

In order to measure the global kinematic PA (i.e. the orientation of line of nodes -- $\Psi_0$)  from the stellar and gas velocity fields we used the kinemetry method \citep{krajnovic05}. 
This method extracts general kinematic properties of the galaxies by the symmetrization of the observed velocity fields, without the need of any assumption on the geometry of the stellar distribution. To obtain the global kinematic PA, the kinemetry method performs the symmetrization of the observed velocity fields. In this process for each possible PA it is created a symmetric velocity field $V'(x,y)$, with the PA oriented along the x axis. The symmetric velocity field is obtained by changing the mean velocity of each bin for the weighted average of the corresponding velocity in the four quadrants of the velocity field. The global kinematic PA is the one that minimizes $\chi^{2}$= $\sum_{n=1}^{N}$($V'(x,y)$ - $V(x,y)$/$\Delta$V)$^{2}$, where $V(x,y)$ is the value of observed velocity field at the position ($x$, $y$).

We used the IDL routine  {\it fit$_-$kinematic$_-$pa.pro}\footnote{This routine was developed by M. Cappellari and is available at \\ http://www-astro.physics.ox.ac.uk/$~$mxc/software}, which is an implementation of the kinemetry method and allows the measurement of the global kinematic PA  and systemic velocity of the galaxy from the observed velocity fields. The routine is an implementation of the  method presented in Appendix C of \citet{krajnovic06} and has been used to study the stellar kinematics of large samples of galaxies, as for example the SAURON \citep{cappellari07} and ATLAS$^{\rm 3D}$ \citep{krajnovic11} surveys.

\section{Results}\label{results}
We have performed measurements for the stellar and gas kinematics and emission-line fluxes for H$\beta$, [O\,{\sc iii}]\,$\lambda$5007, H$\alpha$, [N\,{\sc ii}]\,$\lambda\lambda$6549,83 and [S\,{\sc ii}]\,$\lambda\lambda$6716,31. With the aim of testing our measurements, we have compared the emission-line fluxes, centroid velocities and velocity dispersions with measurements provided by the MaNGA Data Analysis Pipeline (DAP -- Westfall K. B., in prep.), as part of the MPL-7.

 Figure\,\ref{Fig.1} shows an example of our measurements (top row)  compared with those from the DAP (bottom row) for the AGN {\it{mangaid}} 1-339163.  
 The first column shows a map of the continuum emission, the following columns exhibit  maps of emission line fluxes for [O\,{\sc iii}]5007\,\AA, H$\alpha$ and [N\,{\sc ii}]6583\,\AA, respectively. The comparison between the top and bottom rows show that our flux measurements are similar to those provided by the DAP.  
%  This galaxy is an example of how our methodology is similar the measurements of the line and continuum fluxes to larger distances from the nucleus when compared with the results from the DAP. %In almost all galaxies from our sample we were able to increase the spatial coverage of the continuum and emission-line maps. 
%   Our flux measurements agree with those of the DAP within 10\% for all galaxies.
   %, but the Butterworth filter allowed us to obtain reliable measurements of the gas and stellar kinematics at greater distances from the nucleus, as compared to those provided by the MaNGA DAP.
  %When compared to DAP, our measurements present slightly higher peak values in the flux maps (about x\%?), this is caused mainly by ..., and we expect this not (or do) affect ...

In Figure \,\ref{Fig.2} we show the velocity fields for the same galaxy {\it{mangaid}} 1-339163. We present the stellar velocity field together with the gas velocity fields derived for the same emission lines presented in Fig. \ref{Fig.1}. For comparison, we show our results in the top row while the results from DAP are shown in the bottom row. The comparison show that the two velocity fields are similar, although the DAP maps are noisier.

The comparison of the velocity dispersion maps obtained by us and from the DAP is shown in Figure\,\ref{Fig.3}, following the same pattern of organization as the previous figures. As for the centroid velocity and emission-line flux maps, the $\sigma$ maps from DAP are noisier than ours.  The gas and stellar $\sigma$ values will be used to search for outflows in the central region of the galaxies of our sample. 

As noticed in Figs~\ref{Fig.1}--\ref{Fig.3}, our measurements are in general consistent with those provided by DAP, but the spatial filtering of the data allows the exclusion of spurious data, as clearly seen in the maps for the [N\,{\sc ii}]6583\,\AA and H$\alpha$ velocity dispersion, for which the maps constructed using the DAP shows a spurious feature at 4$^{\prime\prime}$ east of the nucleus, which is not present in our measurements. On the other hand, the DAP has the advantage of providing measurements for all emission lines present in the galaxy spectra, while we fit only the strongest lines. However, a detailed comparison of our measurements and those provided by DAP is beyond the scope of this paper.

In order to verify if outflows of gas from the central AGN affects significantly the kinematics of AGN hosts, we can compare the kinematic position angle ($\Psi_0$) of the gas and stellar velocity fields. The motion of the stars is dictated by the gravitational potential of the galaxy, while for the gas, an additional component due to outflows is expected for the AGN. By comparing the difference between the $\Psi_0$ values derived from the gas and stellar velocity fields for AGN and control samples, one should expect larger differences for the AGN if strong outflows are present.% Such outflows would be powerful enough to disturb the gas velocity field and thus larger offsets between gas and stellar $\Psi_0$ would be expected for AGN.  

We derived $\Psi_0$ for the stellar  and gas velocity fields using  [O\,{\sc iii}]5007\,\AA,  H$\alpha$ and [N\,{\sc ii}]6583\,\AA\ emission lines. In Figure\,\ref{Fig.4} we show two examples of the observed and symmetrized velocity fields for two AGN ({\it mangaid} 1-95092 and {\it mangaid} 1-351790). This figure illustrates two distinct results: (i) the $\Psi_0$ from distinct emission-line velocity fields are very similar to each other for both galaxies; (ii) for the galaxy {\it mangaid} 1-95092 the $\Psi_0$ derived from the stellar velocity field is very similar to that  derived for the gas velocity field. (iii) in the case of the AGN host {\it mangaid}  1-351790 the orientation of the kinematic major axis of the stellar and gas velocity fields show a significant offset. In Figure~\ref{Fig.5} we show a similar figure for two control galaxies: {\it mangaid} 12-129446 and {\it mangaid} 1-178838, showing similar results as those observed for the AGN: similar $\Psi_0$ for all emission lines and in one case a distinct $\Psi_0$ for the gas and stars. From Fig. \ref{Fig.4} and Fig. \ref{Fig.5} we can conclude that for this galaxies both AGN and controls present a rotation pattern in the stellar as well as in the gas velocity fields. In Table \ref{Tab.1} we present the kinematic position angle derived for all galaxies of our sample.%It is possible to see that for AGN MaNGAID 1-95092 and the control MaNGAID 12-129446 there is no significant offset between the orientation of the kinematic major axis of the stellar and gas components. On the other hand, for the AGN {\it{mangaid}} 1-351790 and the control  MaNGA ID 1-178838 there are clear kinematic offsets between the stellar and gas velocity fields, respectively, $\Delta$PA=39$^{\circ}$ and $\Delta$PA=41$^{\circ}$ with respect to  [O\,{\sc iii}].

\section{Discussion} \label{discussion}

In order to investigate if the AGN feedback in our sample is powerful enough to disturb the gas kinematics on galactic scales and change the orientation of the kinematic major axis of the galaxy, we calculated the frequency of occurrence of a given PA offset in the AGN and control samples. We computed the difference in the $\Psi_{0 \star}$ of the stellar velocity field with respect to the $\Psi_{\rm 0 gas}$ derived for [O\,{\sc iii}]5007\,\AA, H$\alpha$ and [N\,{\sc ii}]6583\,\AA\ emission-line velocity fields. The resulting histograms of the PA offsets ($\Delta$PA$=|\Psi_{\rm 0 gas} -\Psi_{0 \star}|$) are presented in Figure\,\ref{Fig.6}. The top panels show the results using the [O\,{\sc iii}] velocity fields, while the middle panel show these results for H$\alpha$ and the bottom panel for[N\,{\sc ii}]. AGN are represented by blue colors and control galaxies are shown red.

% I, in order to determine if exist a difference between AGN and non-AGN offsets.
We find no clear difference in the distribution of $\Delta$PA for the AGN and control samples.  Similar values of $\Delta$PA are observed for distinct emission lines. Although a few galaxies display large $\Delta$PA values, for  most of them $\Delta$PA  is smaller than 30$^{\circ}$.  For 79\,\% of AGN and 81\,\% of control galaxies the PA offsets are  smaller than 30$^{\circ}$ as measured using the [O\,{\sc iii}]5007\,\AA\  velocity field as representative of the gas velocity field. This result indicates that the AGN feedback is not strong enough to disturb $-$ more than in a control sample $-$ the gas kinematics on the galactic scales probed by MaNGA. Indeed, the sample of active galaxies used here is composed mainly by low-luminosity AGN \citep{rembold17}, for which outflows from the accretion disk are expected to be weak and thus the gas velocity fields of these AGN hosts on galactic scales are expected to be driven by the gravitational potential of the galaxy. Besides that, \citet{wylezalek17} only find evidence for an AGN-driven outflow in a MaNGA-selected AGN candidate when zoom into the center with higher spatial resolution. The resolution of MaNGA is only 1\farcs5--2\farcs5, so a lot of small scale outflows may be hidden. We do not find any clear difference in the $\Delta$PA  of high and low luminosity AGN.

\citet{penny18} analyzed low-mass galaxies (M$_{\star}$ $\lesssim$ 5$\times$10$^{9}$M$_{\odot}$) of the SDSS-IV MaNGA and found that five galaxies of their sample of 13 possible dwarf AGN host, exhibit ionized gas components in H$\alpha$ that are kinematically offset from their stellar velocity field and these objects have AGN-like emission line ratios at their centers. This fact has been interpreted as due to a recent accretion episode or outflow. Furthermore, \citet{penny18} suggest that AGN feedback may play an important role in these low-mass galaxies. Their sample can be considered an analogous of the ``Red Geysers" galaxies reported by \citet{cheung16} using MaNGA data. These galaxies do not show recent star formation activity, most of them harbor very low luminosity AGN, showing large scale bi-polar outflows in ionized gas and interpreted as being originated by centrally driven winds due to a Radiatively Inefficient Accretion Flow onto the Supermassive Black Hole. These galaxies show stellar and gas kinematic major axes misaligned and 
account for 10\% of the population of galaxies with masses of the order of $2\times10^{10}$\,M$_\odot$ that do not show recent star formation episodes. Although some galaxies of our sample show $\Delta$PA$>30^\circ$, as seen in Fig.\,\ref{Fig.6}, the fraction of AGN  and control galaxies  with significant PA offset are similar (21\,\%  and 19\,\% for AGN and control sample, respectively), suggesting that these offsets are not associated to the presence AGN and probably they are just statistical fluctuations. Thus, we show that standard AGN do not follow the same behavior of ``Red Geyser" galaxies analyzed by \citet{cheung16} and the low-mass galaxies presented in \citet{penny18}, as we do not detect significant PA offsets. 

The fact that there are no significant PA offsets in our sample does not necessarily mean that the AGN do not show outflows, although it implies they do not play an important role in the galaxy scale gas kinematics. However, AGN driven outflows could be seen on  smaller scales.

In order to  search for signatures of outflows closer to the nuclei of the galaxies, we have compared the stellar and gas velocity dispersion values within the inner 2\farcs5 diameter of the galaxies of our sample, as this aperture corresponds to the angular resolution of the MaNGA datacubes. In Table~\ref{Tab.1} we show these velocity dispersion values. On average, the 2\farcs5 aperture  corresponds to a physical scale of $\sim$2\,kpc at the typical redshift of the sample galaxies. In order to quantify the differences between the stellar and gas velocity dispersions measured in the central regions we calculated the parameter $\sigma_{\text{frac}}$, defined as:
\begin{equation}
\sigma_{\text{frac}} = \frac{\sigma_{\text{gas}} - \sigma_{\star}} {\sigma_{\star}},
\end{equation}
which measures the fractional difference between the gas and stellar velocity dispersion, and thus higher values of $\sigma_{\text{frac}}$ are indicative of a disturbed kinematics (not only due to the gravitational potential of the galaxy) and most probably due to outflows.
% Besides that, we have calculated the fractional $\sigma_{\rm frac}$ differences between the stellar velocity dispersion and gas velocity dispersion $\left(\sigma_{\rm frac}=\frac{\sigma_{\rm gas}-\sigma_{\rm stars}}{\sigma_{\rm stars}}\right)$, as well derived the luminosity for [O\,{\sc iii}] in the same place.

%{\color{magenta} GABRIELE (TO DO) Here you could show a figure containing the histograms of $\sigma_{\text{frac, AGN}}$ and $\sigma_{\text{frac,CS}}$, because we know that both distributions are different, and  you will derive the mean value (and std?) for each distribution. Here you could say:}
 We see a trend of AGN having generally higher $\sigma_{\rm frac}$ values than inactive galaxies as can be seen in the distributions shown in Figure \ref{Fig.7}. The median values of $\sigma_{\rm frac}$ for AGN and control sample are  $<\sigma_{\rm frac}>_{\rm AGN}=0.04$ and $<\sigma_{\rm frac}>_{\rm CTR}=-0.23$, respectively. Besides that, we note that 90\% of AGN have $\sigma_{\rm frac}$ larger than $-0.22$ and 75\% of them have values larger than  $-0.13$. For the control sample, 90\% of the galaxies show $\sigma_{\rm frac}<0.12$ and for 75\% of the sample $\sigma_{\rm frac}<-0.04$. The result of the Anderson-Darling statistical test returns a p-value of $10^{-5}$, that confirms  that the AGN and inactive galaxies follow distinct distributions in $\sigma_{\rm frac}$. We thus conclude that the parameter  \sigfrac\  can be used as an indicative of AGN activity.
%{\color{magenta} Now we are going to explore what may be causing this observed behavior correlating $\sigma_{\text{frac}}$ with $L_{[{O}{iii}]}$.}

We derived the luminosity of the [{O}{\sc iii}]$\lambda5007$\,\AA\ emission line ($L_{[{O}{iii}]}$)  of each galaxy (Table \ref{Tab.1}) using the flux measurements obtained with the {\sc gandalf} code within the same aperture used to measure the  $\sigma_{\rm frac}$, and then investigated a possible correlation between $\sigma_{\text{frac}}$ and   $L_{[{O}{\sc iii}]}$. Figure \ref{Fig.8} shows the plot of $L_{[{O}{\sc iii}]}$ vs. $\sigma_{\rm frac}$ for the AGN and control samples. There is a clear positive correlation between  $\sigma_{\text{frac}}$ and  $L_{[{O}{\sc iii}]}$, with a Spearman correlation coefficient of 0.53 and a p-value of 10$^{-14}$.  However, it should be noticed that the observed correlation could be artificially produced, as the AGN and inactive galaxies clearly show distinct distributions in $\sigma_{\rm frac}$ (Fig.~\ref{Fig.7}). The Spermann test returns a p-value of 0.06 for the AGN sample and  10$^{-5}$ for the control sample, meaning that no strong correlation is found between the $L_{[{O}{iii}]}$ and $\sigma_{\text{frac}}$ for the AGN sample alone, while these parameters are correlated for the control sample. The absence of correlation for the AGN sample may be due to the fact that our sample covers only a small range of luminosities, as most objects are low-luminosity AGN \citep{rembold17}.  Fig.~\ref{Fig.7}  shows  a trend of AGN having higher $\sigma_{\rm frac}$ values than inactive galaxies. The same  trend can also be observed in Fig.~\ref{Fig.8}. This result  can be interpreted as the higher values seen for AGN as compared to control galaxies being due to winds originated in the AGN.  Thus, although the AGN of the sample do not show powerful outflows that can affect the gas kinematics on galactic scales, they do show small scale outflows (within the inner 1-2 kpc).

Our results can be compared with those obtained from single aperture spectra. For example, \citet{woo17} find that there is a trend of the [O\,{\sc iii}]5007\,\AA\ velocity dispersion to increase with the increase of the AGN luminosity in a sample of $\sim$110,000 AGN and star-forming (SF) galaxies at $z<$0.3. This trend is also present in composite objects and is not clear for star-forming galaxies. They interpreted this result as due to strong gas outflows in high luminosity AGN, indicating that AGN energetics are driving these outflows. They find also lower average [O\,{\sc iii}] velocity dispersion values for star-forming galaxies.  Our result is in good agreement with theirs. In addition, optical observations \citep{wylezalek16}, radio observations \citep{zakamska14} and molecular gas \citep{veilleux13} as well as theoretical models \citep{zubovas12b} have suggested that the AGN needs to have enough luminosity for the gas to be pushed out of the galactic potential. This is in agreement with our results, where we see a positive correlation between $\sigma_{\text{frac}}$ and luminosity.
%, and has the advantage that our AGN sample of AGN is compared to the inactive galaxy sample, matched to the AGN hosts  properties and thus excluding any possible selection bias. 

\section{Conclusions} \label{conclusions}

We have mapped the gas and stellar kinematics of a sample of 62 AGN and 109 control galaxies (inactive galaxies) in order to investigate the effect of the AGN in the large and small scale gas kinematics of the AGN host galaxies. 
We detect evidence of nuclear gas outflows in the 62 AGN, but conclude they are not powerful enough to play an important role in the gas kinematics on galactic scales.
The main conclusions of our work are:

\begin{itemize}
%\item Our measurements for emission-line fluxes, stellar and gas kinematics done with GANDALF agree with those available in MaNGA-DAP. But for some galaxies, we have obtained a more extended fit in places away from the center of galaxy than DAP. Besides that the peak of emission is more intense in our flux maps than in DAP, our velocity and velocity dispersion fields are smoother.

\item There is no significant difference in the $\Delta$PA between active and inactive galaxies, indicating that the galaxy scale gas kinematics is dominated by orbital motion in the gravitational potential of the galaxies, instead of outflows driven by the central AGN.

\item We found that the difference between the orientation of the kinematic major axes of the gas and stars ($\Delta$PA) is larger than 30$^\circ$ for 13 (21\,\%) AGN and 21 control galaxies (19\,\%) using the [O\,{\sc iii}]5007\,\AA\  kinematics.

 \item The AGN show larger fractional differences in the velocity dispersions of the gas and stars $\sigma_{\rm frac}=\frac{\sigma_{OIII}-\sigma_{\star}}{\sigma_\star}$ than inactive galaxies within the inner 2\farcs5 diameter, that corresponds to 1-2kpc at the galaxies. The mean values are $\sigma_{\rm frac}$=0.05 for the AGN  and $\sigma_{\rm frac}$=-0.23 for the control sample. This difference is interpreted as being due to outflows from the active nuclei.  This indicates that, although the AGN of our sample do not affect the gas kinematics on large scale, it does affect it at least within the inner kpc. 
\item A correlation between the [O\,{\sc iii}]5007\,\AA\ luminosity and $\sigma_{\rm frac}$ is observed when putting together the AGN and control samples.%We find a correlation between [O\,{\sc iii}]5007\,\AA\ luminosity and $\sigma_{frac}$ when putting together the AGN and control samples, indicating the the impact of the outflows increase with the luminosity of the AGN.

\end{itemize}

\section*{Acknowledgements}

Funding for the Sloan Digital Sky Survey IV has been provided by the Alfred P. Sloan Foundation and the Participating Institutions.  SDSS-IV acknowledges
support and resources from the Center for High-Performance Computing at the University of Utah. The SDSS web site is www.sdss.org. SDSS-IV is managed by the
Astrophysical Research Consortium for the Participating Institutions of the SDSS Collaboration including the Brazilian Participation Group, the Carnegie Institution for Science,
Carnegie  Mellon  University,  the  Chilean  Participation  Group,  Harvard-Smithsonian Center for Astrophysics, Instituto de Astrof\'isica de Canarias, The Johns Hopkins University, Kavli Institute for the Physics and Mathematics of the Universe (IPMU) / University of Tokyo, Lawrence Berkeley National Laboratory, Leibniz Institut f\"ur Astro
physik  Potsdam  (AIP),  Max-Planck-Institut  f\"ur  Astrophysik  (MPA Garching),  Max-Planck-Institut f\"ur Extraterrestrische Physik (MPE), Max-Planck-Institut f\"ur Astronomie
(MPIA Heidelberg), National Astronomical Observatory of China, New Mexico State University, New York University, The Ohio State University, Pennsylvania State University, 
Shanghai  Astronomical  Observatory,  United  Kingdom  Participation  Group, Universidad Nacional Aut\'onoma de M\'exico, University of Arizona, University of Colorado Boulder,
University of Portsmouth, University of Utah, University of Washington, University of Wisconsin, Vanderbilt University, and Yale University.
We thank the support of the Instituto Nacional de Ci\^encia
e Tecnologia (INCT) e-Universe (CNPq grant 465376/2014-2). 

This study was financed in part by the Coordena\c c\~ao de
Aperfei\c coamento de Pessoal de N\'ivel Superior - Brasil (CAPES) -
Finance Code 001, Conselho Nacional de Desenvolvimento Cient\'ifico e Tecnol\'ogico (CNPq) and Funda\c c\~ao de Amparo \`a pesquisa do Estado do RS (FAPERGS).

%%%%%%%%%%%%%%%%%%%%%%%%%%%%%%%%%%%%%%%%%%%%%%%%%%



%\bibliographystyle{mnras}


%%%%%%%%%%%%%%%%%%%%%%%%%%%%%%%%%%%%%%%%%%%%%%%%%%

%%%%%%%%%%%%%%%%% APPENDICES %%%%%%%%%%%%%%%%%%%%%

\appendix
\section{Kinematic measurements of the AGN and control samples}

%%%%%%%%%%%%%%%%%%%%%%%%%%%%%%%%%%%%%%%%%%%%%%%%%%

\begin{table*}
\caption{Kinematic measurements of the AGN and control samples. Col. 1: {\it mangaid}; Col. 2: Logarithm of luminosity of [O\,{\sc iii}]5007\,\AA\ in erg\,s$^{-1}$ within the inner 2\farcs5 diameter; velocity dispersion in km s$^{-1}$ obtained for stars, [O\,{\sc iii}]5007\,\AA, H$\alpha$, [N\,{\sc ii}]6583\,\AA\ (cols. 3-6), measured within the inner 2\farcs5 diameter, and  kinematic position angles ($\Psi_0$) for the stars,  [O\,{\sc iii}]5007\,\AA, H$\alpha$, [N\,{\sc ii}]6583\,\AA\ (cols. 7-10) . In each block we have the parameters obtained for AGN (first line) and their controls (second and third lines).}
\begin{tabular}{cccccccccc}
\toprule
{\it mangaid} & log$_{10}L_{[{O}{iii}]}$&  $\sigma_{\star}$ &  $\sigma_{[{O}{iii}]}$  &  $\sigma_{H\alpha}$ & $\sigma_{[{N}{ii}]}$ & $\Psi_0$$_{\star}$ & $\Psi_0$ [O\,{\sc iii}]  & $\Psi_0$ H$\alpha$ & $\Psi_0$ [N\,{\sc ii}]  \\
\midrule
1-558912 & 41.32 & 269.0$\pm$3.0 & 231.0$\pm$1.0 & 204.0$\pm$2.0 & 209.0$\pm$2.0 & 148.0$\pm$0.5 & 165.0$\pm$0.5 & 117.5$\pm$0.5 & 158.5$\pm$0.5 \\
1-71481 & 39.38 & 284.0$\pm$2.0 & 194.0$\pm$12.0 & 275.0$\pm$12.0 & 219.0$\pm$11.0 & 174.0$\pm$0.5 & 146.5$\pm$0.5 & 9.5$\pm$0.5 & 163.5$\pm$0.5 \\
1-72928 & 39.45 & 253.0$\pm$1.0 & 268.0$\pm$19.0 & 161.0$\pm$20.0 & 295.0$\pm$126.0 & 138.0$\pm$0.5 & 134.0$\pm$0.5 & 121.0$\pm$0.5 & 137.5$\pm$0.5 \\
\\
1-269632 & 41.35 & 153.0$\pm$1.0 & 202.0$\pm$2.0 & 137.0$\pm$2.0 & 155.0$\pm$1.0 & 16.0$\pm$2.2 & 173.5$\pm$0.5 & 11.5$\pm$2.0 & 14.0$\pm$0.8 \\
1-210700 & 39.7 & 199.0$\pm$2.0 & 150.0$\pm$4.0 & 112.0$\pm$1.0 & 131.0$\pm$1.0 & 130.0$\pm$0.5 & 137.5$\pm$0.5 & 134.5$\pm$0.5 & 119.5$\pm$0.5 \\
1-378795 & 39.4 & 172.0$\pm$2.0 & 148.0$\pm$5.0 & 109.0$\pm$1.0 & 128.0$\pm$3.0 & 19.5$\pm$0.5 & 25.5$\pm$0.5 & 20.5$\pm$0.5 & 19.0$\pm$0.5 \\
\\
1-258599 & 41.87 & 349.0$\pm$3.0 & 208.0$\pm$1.0 & 170.0$\pm$1.0 & 166.0$\pm$1.0 & 81.5$\pm$0.5 & 98.5$\pm$0.5 & 122.5$\pm$0.5 & 108.5$\pm$0.5 \\
1-93876 & 39.11 & 223.0$\pm$1.0 & 230.0$\pm$20.0 & 275.0$\pm$18.0 & 98.0$\pm$18.0 & 146.0$\pm$0.5 & 148.5$\pm$0.5 & 149.5$\pm$0.5 & 136.0$\pm$0.5 \\
1-166691 & 38.76 & 231.0$\pm$1.0 & 281.0$\pm$25.0 & 175.0$\pm$15.0 & 221.0$\pm$39.0 & 92.0$\pm$0.5 & 146.5$\pm$0.5 & 34.5$\pm$0.5 & 95.0$\pm$0.5 \\
\\
1-72322 & 41.25 & 287.0$\pm$3.0 & 156.0$\pm$2.0 & 170.0$\pm$2.0 & 182.0$\pm$3.0 & 107.0$\pm$0.8 & 114.5$\pm$0.5 & 116.0$\pm$0.5 & 118.5$\pm$0.5 \\
1-121717 & 39.86 & 223.0$\pm$2.0 & 334.0$\pm$26.0 & 167.0$\pm$3.0 & 173.0$\pm$4.0 & 134.0$\pm$0.5 & 111.0$\pm$0.5 & 134.0$\pm$0.5 & 134.0$\pm$0.5 \\
1-43721 & 39.48 & 250.0$\pm$1.0 & 265.0$\pm$14.0 & 132.0$\pm$2.0 & 140.0$\pm$2.0 & 55.0$\pm$2.2 & 52.0$\pm$0.5 & 51.5$\pm$0.5 & 43.0$\pm$0.5 \\
\\
1-121532 & 40.55 & 260.0$\pm$1.0 & 318.0$\pm$10.0 & 280.0$\pm$8.0 & 300.0$\pm$8.0 & 101.0$\pm$0.5 & 148.5$\pm$0.5 & 10.0$\pm$0.5 & 77.0$\pm$0.5 \\
1-218427 & 38.9 & 386.0$\pm$13.0 & 130.0$\pm$21.0 & 616.0$\pm$191.0 & 466.0$\pm$133.0 & 136.5$\pm$0.5 & 178.0$\pm$0.5 & 136.5$\pm$0.5 & 136.0$\pm$0.5 \\
1-177493 & 39.91 & 172.0$\pm$2.0 & 145.0$\pm$2.0 & 156.0$\pm$3.0 & 111.0$\pm$2.0 & 142.0$\pm$3.8 & 145.0$\pm$2.0 & 158.0$\pm$0.5 & 149.0$\pm$0.5 \\
\\
1-209980 & 40.66 & 179.0$\pm$3.0 & 218.0$\pm$1.0 & 233.0$\pm$1.0 & 245.0$\pm$1.0 & 50.0$\pm$0.8 & 52.0$\pm$0.5 & 55.0$\pm$0.5 & 51.5$\pm$0.5 \\
1-295095 & 38.76 & 101.0$\pm$1.0 & 112.0$\pm$3.0 & 80.0$\pm$1.0 & 88.0$\pm$2.0 & 1.0$\pm$2.2 & 3.0$\pm$0.8 & 178.0$\pm$1.5 & 1.0$\pm$1.2 \\
1-92626 & 39.41 & 181.0$\pm$2.0 & 116.0$\pm$2.0 & 138.0$\pm$2.0 & 183.0$\pm$10.0 & 10.5$\pm$0.8 & 13.5$\pm$0.5 & 13.0$\pm$0.5 & 13.0$\pm$0.5 \\
\\
1-44379 & 40.49 & 133.0$\pm$2.0 & 168.0$\pm$2.0 & 126.0$\pm$3.0 & 159.0$\pm$3.0 & 15.0$\pm$0.5 & 14.0$\pm$0.5 & 16.5$\pm$0.5 & 16.5$\pm$0.8 \\
1-211082 & 38.86 & 153.0$\pm$1.0 & 89.0$\pm$3.0 & 118.0$\pm$2.0 & 158.0$\pm$1.0 & 157.5$\pm$0.5 & 163.5$\pm$0.5 & 159.0$\pm$0.5 & 159.5$\pm$0.5 \\
1-135371 & 38.9 & 144.0$\pm$2.0 & 105.0$\pm$4.0 & 116.0$\pm$2.0 & 147.0$\pm$4.0 & 67.5$\pm$0.8 & 72.5$\pm$0.5 & 66.5$\pm$0.8 & 66.5$\pm$0.5 \\
\\
1-149211 & 40.5 & 107.0$\pm$1.0 & 116.0$\pm$0.0 & 121.0$\pm$1.0 & 134.0$\pm$2.0 & 41.0$\pm$0.5 & 166.0$\pm$0.5 & 174.5$\pm$0.5 & 25.5$\pm$0.5 \\
1-377321 & 40.15 & 182.0$\pm$6.0 & 77.0$\pm$0.0 & 77.0$\pm$0.0 & 80.0$\pm$0.0 & 17.5$\pm$0.5 & 26.5$\pm$0.5 & 44.0$\pm$0.5 & 85.5$\pm$0.5 \\
1-491233 & 39.09 & 106.0$\pm$1.0 & 94.0$\pm$1.0 & 83.0$\pm$1.0 & 87.0$\pm$1.0 & 120.5$\pm$1.5 & 105.0$\pm$1.8 & 118.5$\pm$1.2 & 118.5$\pm$1.5 \\
\\
1-173958 & 40.49 & 101.0$\pm$1.0 & 207.0$\pm$2.0 & 114.0$\pm$1.0 & 142.0$\pm$2.0 & 10.0$\pm$4.2 & 32.0$\pm$0.5 & 18.5$\pm$3.2 & 21.0$\pm$0.5 \\
1-247456 & 39.3 & 879.0$\pm$34.0 & 172.0$\pm$13.0 & 1090.0$\pm$327.0 & 749.0$\pm$91.0 & 149.0$\pm$1.2 & 169.0$\pm$0.5 & 167.5$\pm$0.8 & 167.5$\pm$0.5 \\
1-24246 & 39.1 & 110.0$\pm$1.0 & 134.0$\pm$4.0 & 94.0$\pm$2.0 & 105.0$\pm$2.0 & 139.0$\pm$1.5 & 134.0$\pm$0.8 & 134.0$\pm$1.0 & 121.5$\pm$0.5 \\
\\
1-338922 & 40.39 & 284.0$\pm$2.0 & 215.0$\pm$5.0 & 215.0$\pm$3.0 & 212.0$\pm$1.0 & 47.5$\pm$3.8 & 110.5$\pm$0.5 & 103.0$\pm$0.5 & 129.5$\pm$0.5 \\
1-286804 & 39.81 & 179.0$\pm$3.0 & 204.0$\pm$5.0 & 171.0$\pm$4.0 & 173.0$\pm$3.0 & 168.5$\pm$0.5 & 164.0$\pm$1.2 & 169.5$\pm$0.5 & 153.0$\pm$0.5 \\
1-109493 & 39.56 & 257.0$\pm$2.0 & 278.0$\pm$17.0 & 521.0$\pm$140.0 & 215.0$\pm$8.0 & 102.5$\pm$0.5 & 42.0$\pm$0.5 & 58.5$\pm$0.5 & 137.5$\pm$0.5 \\
\\
1-279147 & 40.29 & 116.0$\pm$1.0 & 163.0$\pm$1.0 & 101.0$\pm$1.0 & 122.0$\pm$1.0 & 31.5$\pm$4.2 & 51.5$\pm$0.5 & 54.0$\pm$0.5 & 1.0$\pm$0.5 \\
1-283246 & 38.68 & 142.0$\pm$0.0 & 126.0$\pm$3.0 & 157.0$\pm$8.0 & 140.0$\pm$3.0 & 37.0$\pm$2.8 & 59.0$\pm$1.0 & 29.5$\pm$0.5 & 31.0$\pm$0.5 \\
1-351538 & 39.26 & 109.0$\pm$1.0 & 125.0$\pm$3.0 & 69.0$\pm$0.0 & 76.0$\pm$1.0 & 160.5$\pm$16.2 & 141.0$\pm$0.5 & 169.0$\pm$1.2 & 176.5$\pm$1.2 \\
\\
1-460812 & 40.48 & 210.0$\pm$3.0 & 179.0$\pm$2.0 & 212.0$\pm$1.0 & 218.0$\pm$3.0 & 78.0$\pm$0.8 & 78.5$\pm$0.5 & 69.5$\pm$0.5 & 68.5$\pm$0.5 \\
1-270160 & 39.63 & 319.0$\pm$3.0 & 331.0$\pm$9.0 & 362.0$\pm$17.0 & 311.0$\pm$10.0 & 111.0$\pm$0.5 & 112.5$\pm$0.5 & 104.0$\pm$0.5 & 116.5$\pm$1.0 \\
1-258455 & 39.15 & 214.0$\pm$1.0 & 184.0$\pm$7.0 & 184.0$\pm$3.0 & 194.0$\pm$7.0 & 131.0$\pm$1.5 & 130.0$\pm$0.5 & 128.5$\pm$0.5 & 127.5$\pm$0.5 \\
\\
1-92866 & 40.3 & 241.0$\pm$1.0 & 244.0$\pm$2.0 & 217.0$\pm$1.0 & 258.0$\pm$2.0 & 18.5$\pm$1.5 & 126.5$\pm$0.5 & 151.0$\pm$0.5 & 123.5$\pm$0.5 \\
1-94514 & 38.69 & 237.0$\pm$1.0 & 229.0$\pm$18.0 & 373.0$\pm$71.0 & 386.0$\pm$23.0 & 147.5$\pm$1.0 & 147.0$\pm$0.5 & 143.0$\pm$0.5 & 135.0$\pm$0.5 \\
1-210614 & 38.91 & 223.0$\pm$1.0 & 152.0$\pm$3.0 & 178.0$\pm$2.0 & 157.0$\pm$3.0 & 131.0$\pm$2.2 & 90.0$\pm$0.5 & 95.0$\pm$0.5 & 82.0$\pm$0.5 \\
\\
1-94784 & 40.48 & 135.0$\pm$1.0 & 142.0$\pm$2.0 & 156.0$\pm$1.0 & 153.0$\pm$2.0 & 65.5$\pm$1.0 & 71.0$\pm$0.5 & 62.5$\pm$0.8 & 67.5$\pm$1.5 \\
1-211063 & 38.75 & 136.0$\pm$1.0 & 76.0$\pm$2.0 & 141.0$\pm$2.0 & 111.0$\pm$2.0 & 170.0$\pm$1.8 & 164.0$\pm$0.5 & 166.5$\pm$1.8 & 165.0$\pm$1.2 \\
1-135502 & 39.16 & 164.0$\pm$1.0 & 117.0$\pm$1.0 & 177.0$\pm$2.0 & 163.0$\pm$1.0 & 98.5$\pm$0.5 & 98.5$\pm$0.5 & 101.5$\pm$0.5 & 102.0$\pm$0.5 \\
\\
1-44303 & 40.15 & 114.0$\pm$1.0 & 135.0$\pm$1.0 & 120.0$\pm$2.0 & 134.0$\pm$2.0 & 56.0$\pm$7.2 & 39.0$\pm$7.8 & 67.5$\pm$2.0 & 62.0$\pm$3.0 \\
1-339028 & 39.22 & 183.0$\pm$1.0 & 133.0$\pm$2.0 & 170.0$\pm$3.0 & 145.0$\pm$1.0 & 179.5$\pm$2.7 & 173.5$\pm$1.0 & 5.5$\pm$0.8 & 9.0$\pm$1.8 \\
1-379087 & 39.27 & 131.0$\pm$1.0 & 104.0$\pm$2.0 & 123.0$\pm$1.0 & 125.0$\pm$1.0 & 39.5$\pm$2.2 & 31.5$\pm$0.5 & 42.5$\pm$0.5 & 44.0$\pm$1.0 \\
\\
1-339094 & 40.2 & 133.0$\pm$1.0 & 160.0$\pm$1.0 & 152.0$\pm$2.0 & 157.0$\pm$2.0 & 0.0$\pm$1.5 & 176.0$\pm$1.8 & 175.0$\pm$0.8 & 8.0$\pm$0.5 \\
1-274646 & 38.91 & 121.0$\pm$1.0 & 82.0$\pm$1.0 & 108.0$\pm$1.0 & 108.0$\pm$1.0 & 61.5$\pm$17.2 & 175.0$\pm$0.5 & 41.0$\pm$0.5 & 22.5$\pm$0.5 \\
1-24099 & 38.34 & 125.0$\pm$2.0 & 84.0$\pm$2.0 & 131.0$\pm$17.0 & 153.0$\pm$6.0 & 125.5$\pm$1.2 & 135.5$\pm$0.8 & 134.5$\pm$0.5 & 121.5$\pm$0.5 \\
\\
\end{tabular}
\label{Tab.1}
\end{table*}

\begin{table*}
\contcaption{}
\begin{tabular}{cccccccccc}
\toprule
{\it mangaid} & log$_{10}L_{[{O}{iii}]}$&  $\sigma_{\star}$ &  $\sigma_{[{O}{iii}]}$  &  $\sigma_{H\alpha}$ & $\sigma_{[{N}{ii}]}$ & $\Psi_0$$_{\star}$ & $\Psi_0$ [O\,{\sc iii}]  & $\Psi_0$ H$\alpha$ & $\Psi_0$ [N\,{\sc ii}]  \\
\midrule
1-137883 & 40.15 & 185.0$\pm$5.0 & 164.0$\pm$3.0 & 133.0$\pm$3.0 & 130.0$\pm$2.0 & 55.5$\pm$2.0 & 56.5$\pm$0.8 & 48.0$\pm$1.2 & 41.5$\pm$0.5 \\
1-178838 & 38.69 & 114.0$\pm$2.0 & 90.0$\pm$5.0 & 107.0$\pm$3.0 & 103.0$\pm$2.0 & 84.5$\pm$1.5 & 44.0$\pm$0.5 & 41.0$\pm$4.8 & 36.0$\pm$2.2 \\
1-36878 & 39.07 & 138.0$\pm$3.0 & 101.0$\pm$1.0 & 90.0$\pm$1.0 & 98.0$\pm$1.0 & 46.0$\pm$1.0 & 26.0$\pm$0.5 & 43.0$\pm$0.5 & 45.5$\pm$1.2 \\
\\
1-48116 & 40.12 & 151.0$\pm$1.0 & 178.0$\pm$1.0 & 127.0$\pm$1.0 & 134.0$\pm$1.0 & 56.0$\pm$1.0 & 65.0$\pm$0.5 & 57.0$\pm$1.0 & 56.0$\pm$0.5 \\
1-386452 & 39.09 & 123.0$\pm$1.0 & 90.0$\pm$1.0 & 89.0$\pm$1.0 & 91.0$\pm$1.0 & 135.0$\pm$2.0 & 131.0$\pm$0.5 & 131.5$\pm$1.0 & 131.0$\pm$1.2 \\
1-24416 & 38.76 & 157.0$\pm$1.0 & 104.0$\pm$1.0 & 162.0$\pm$3.0 & 148.0$\pm$2.0 & 85.0$\pm$0.8 & 79.5$\pm$0.8 & 83.5$\pm$0.5 & 82.5$\pm$0.5 \\
\\
1-256446 & 40.12 & 214.0$\pm$2.0 & 244.0$\pm$1.0 & 231.0$\pm$1.0 & 232.0$\pm$1.0 & 158.0$\pm$11.0 & 62.0$\pm$0.5 & 48.0$\pm$4.2 & 49.5$\pm$0.5 \\
1-322671 & 38.7 & 173.0$\pm$1.0 & 134.0$\pm$12.0 & 360.0$\pm$121.0 & 167.0$\pm$12.0 & 130.0$\pm$3.2 & 109.0$\pm$0.5 & 101.5$\pm$0.5 & 143.0$\pm$0.5 \\
1-256465 & 39.38 & 217.0$\pm$1.0 & 227.0$\pm$4.0 & 187.0$\pm$3.0 & 216.0$\pm$6.0 & 134.0$\pm$16.0 & 75.0$\pm$0.5 & 62.5$\pm$0.5 & 64.0$\pm$0.5 \\
\\
1-95585 & 40.11 & 171.0$\pm$1.0 & 214.0$\pm$3.0 & 211.0$\pm$3.0 & 206.0$\pm$2.0 & 69.5$\pm$1.0 & 62.5$\pm$0.5 & 68.0$\pm$0.5 & 69.5$\pm$0.5 \\
1-166947 & 38.79 & 149.0$\pm$1.0 & 83.0$\pm$3.0 & 143.0$\pm$3.0 & 116.0$\pm$2.0 & 36.0$\pm$3.2 & 123.0$\pm$0.5 & 34.0$\pm$0.8 & 35.5$\pm$0.8 \\
1-210593 & 39.04 & 160.0$\pm$1.0 & 128.0$\pm$3.0 & 146.0$\pm$4.0 & 147.0$\pm$4.0 & 101.5$\pm$1.0 & 107.5$\pm$0.5 & 104.0$\pm$0.8 & 103.0$\pm$0.5 \\
\\
1-135641 & 40.04 & 165.0$\pm$4.0 & 168.0$\pm$2.0 & 155.0$\pm$3.0 & 148.0$\pm$1.0 & 164.5$\pm$0.5 & 168.5$\pm$0.5 & 163.5$\pm$0.5 & 166.5$\pm$0.5 \\
1-635503 & 39.05 & 121.0$\pm$1.0 & 72.0$\pm$1.0 & 82.0$\pm$1.0 & 83.0$\pm$1.0 & 140.0$\pm$1.2 & 145.0$\pm$0.5 & 141.5$\pm$0.5 & 138.5$\pm$0.5 \\
1-235398 & 38.74 & 154.0$\pm$1.0 & 86.0$\pm$2.0 & 123.0$\pm$1.0 & 120.0$\pm$1.0 & 98.0$\pm$0.5 & 102.0$\pm$0.5 & 99.5$\pm$0.5 & 100.0$\pm$0.5 \\
\\
1-259142 & 40.09 & 211.0$\pm$1.0 & 272.0$\pm$5.0 & 248.0$\pm$4.0 & 274.0$\pm$4.0 & 102.5$\pm$0.5 & 100.0$\pm$0.5 & 106.5$\pm$0.5 & 101.5$\pm$0.5 \\
1-55572 & 38.93 & 194.0$\pm$2.0 & 117.0$\pm$1.0 & 208.0$\pm$4.0 & 198.0$\pm$5.0 & 51.0$\pm$0.5 & 52.0$\pm$0.5 & 50.5$\pm$0.5 & 52.0$\pm$0.8 \\
1-489649 & 39.05 & 170.0$\pm$1.0 & 100.0$\pm$2.0 & 170.0$\pm$6.0 & 155.0$\pm$2.0 & 170.5$\pm$1.5 & 169.0$\pm$0.5 & 1.0$\pm$0.5 & 171.5$\pm$0.8 \\
\\
1-109056 & 40.07 & 124.0$\pm$1.0 & 104.0$\pm$1.0 & 104.0$\pm$1.0 & 115.0$\pm$1.0 & 157.0$\pm$1.5 & 159.0$\pm$0.5 & 156.0$\pm$1.0 & 154.0$\pm$1.0 \\
1-73005 & 38.81 & 107.0$\pm$1.0 & 91.0$\pm$4.0 & 91.0$\pm$1.0 & 91.0$\pm$1.0 & 43.5$\pm$2.2 & 59.5$\pm$0.5 & 44.0$\pm$1.0 & 44.0$\pm$1.2 \\
1-43009 & 38.65 & 94.0$\pm$1.0 & 79.0$\pm$2.0 & 83.0$\pm$1.0 & 85.0$\pm$1.0 & 168.0$\pm$1.2 & 172.5$\pm$1.0 & 171.0$\pm$1.0 & 171.5$\pm$1.0 \\
\\
1-24148 & 40.06 & 149.0$\pm$1.0 & 126.0$\pm$1.0 & 172.0$\pm$1.0 & 159.0$\pm$1.0 & 155.5$\pm$0.8 & 150.5$\pm$0.5 & 163.5$\pm$0.5 & 134.5$\pm$0.5 \\
1-285031 & 38.95 & 117.0$\pm$1.0 & 67.0$\pm$1.0 & 83.0$\pm$0.0 & 82.0$\pm$0.0 & 38.0$\pm$0.8 & 38.0$\pm$0.5 & 37.0$\pm$0.8 & 36.5$\pm$0.8 \\
1-236099 & 38.51 & 85.0$\pm$1.0 & 96.0$\pm$3.0 & 102.0$\pm$4.0 & 83.0$\pm$4.0 & 1.0$\pm$1.7 & 138.0$\pm$2.2 & 37.5$\pm$0.5 & 118.5$\pm$0.5 \\
\\
1-166919 & 39.63 & 164.0$\pm$2.0 & 158.0$\pm$3.0 & 153.0$\pm$4.0 & 223.0$\pm$4.0 & 105.0$\pm$1.0 & 106.0$\pm$0.5 & 103.0$\pm$0.5 & 102.5$\pm$0.8 \\
12-129446 & 39.24 & 115.0$\pm$1.0 & 113.0$\pm$3.0 & 84.0$\pm$1.0 & 97.0$\pm$2.0 & 159.0$\pm$1.5 & 161.0$\pm$0.5 & 157.0$\pm$1.0 & 157.5$\pm$0.8 \\
1-90849 & 39.1 & 115.0$\pm$1.0 & 88.0$\pm$2.0 & 81.0$\pm$1.0 & 95.0$\pm$1.0 & 58.5$\pm$1.0 & 51.5$\pm$0.5 & 55.5$\pm$0.5 & 57.5$\pm$0.5 \\
\\
1-248389 & 39.91 & 148.0$\pm$0.0 & 144.0$\pm$1.0 & 175.0$\pm$1.0 & 160.0$\pm$1.0 & 177.0$\pm$1.5 & 145.0$\pm$0.5 & 128.5$\pm$0.5 & 145.5$\pm$0.5 \\
1-94554 & 38.98 & 136.0$\pm$1.0 & 111.0$\pm$3.0 & 135.0$\pm$2.0 & 146.0$\pm$5.0 & 29.5$\pm$2.0 & 37.5$\pm$0.5 & 8.0$\pm$0.8 & 44.0$\pm$0.5 \\
1-245774 & 38.97 & 130.0$\pm$1.0 & 99.0$\pm$2.0 & 98.0$\pm$1.0 & 107.0$\pm$1.0 & 85.5$\pm$1.2 & 91.5$\pm$0.5 & 85.5$\pm$0.8 & 87.0$\pm$0.5 \\
\\
1-321739 & 39.9 & 172.0$\pm$4.0 & 145.0$\pm$2.0 & 109.0$\pm$1.0 & 123.0$\pm$1.0 & 172.0$\pm$0.5 & 163.5$\pm$0.5 & 174.0$\pm$0.5 & 179.0$\pm$0.5 \\
1-247417 & 38.77 & 107.0$\pm$1.0 & 82.0$\pm$2.0 & 85.0$\pm$1.0 & 90.0$\pm$1.0 & 142.0$\pm$0.5 & 141.0$\pm$0.5 & 141.0$\pm$0.5 & 141.5$\pm$0.5 \\
1-633994 & 38.79 & 238.0$\pm$26.0 & 99.0$\pm$2.0 & 131.0$\pm$4.0 & 201.0$\pm$21.0 & 124.5$\pm$0.8 & 135.5$\pm$0.5 & 123.0$\pm$1.0 & 129.5$\pm$0.5 \\
\\
1-234618 & 39.99 & 182.0$\pm$17.0 & 167.0$\pm$3.0 & 123.0$\pm$9.0 & 137.0$\pm$1.0 & 23.0$\pm$0.5 & 16.0$\pm$1.2 & 19.0$\pm$0.5 & 23.0$\pm$0.8 \\
1-282144 & 39.16 & 83.0$\pm$1.0 & 93.0$\pm$2.0 & 70.0$\pm$1.0 & 75.0$\pm$1.0 & 179.0$\pm$2.5 & 163.0$\pm$0.5 & 179.0$\pm$1.5 & 165.5$\pm$0.5 \\
1-339125 & 38.97 & 181.0$\pm$2.0 & 132.0$\pm$3.0 & 284.0$\pm$44.0 & 653.0$\pm$231.0 & 170.5$\pm$1.5 & 179.5$\pm$0.5 & 141.0$\pm$0.5 & 165.0$\pm$0.5 \\
\\
1-229010 & 39.88 & 224.0$\pm$1.0 & 237.0$\pm$2.0 & 268.0$\pm$4.0 & 228.0$\pm$1.0 & 177.5$\pm$0.5 & 1.5$\pm$0.5 & 178.0$\pm$0.5 & 0.5$\pm$0.5 \\
1-210962 & 39.01 & 194.0$\pm$1.0 & 117.0$\pm$2.0 & 147.0$\pm$2.0 & 157.0$\pm$2.0 & 91.5$\pm$1.2 & 96.5$\pm$0.5 & 89.0$\pm$0.5 & 89.5$\pm$0.5 \\
1-613211 & 38.65 & 242.0$\pm$1.0 & 108.0$\pm$3.0 & 192.0$\pm$3.0 & 188.0$\pm$4.0 & 22.0$\pm$0.5 & 16.5$\pm$0.5 & 18.0$\pm$0.5 & 22.5$\pm$0.5 \\
\\
1-211311 & 39.83 & 114.0$\pm$1.0 & 173.0$\pm$2.0 & 191.0$\pm$2.0 & 219.0$\pm$3.0 & 160.0$\pm$3.0 & 156.0$\pm$0.5 & 167.5$\pm$2.5 & 166.0$\pm$2.0 \\
1-25688 & 38.35 & 110.0$\pm$1.0 & 64.0$\pm$2.0 & 110.0$\pm$2.0 & 115.0$\pm$1.0 & 20.5$\pm$1.5 & 13.5$\pm$0.5 & 17.0$\pm$0.8 & 19.5$\pm$1.5 \\
1-94422 & 38.87 & 129.0$\pm$1.0 & 82.0$\pm$1.0 & 131.0$\pm$2.0 & 123.0$\pm$1.0 & 47.0$\pm$1.0 & 28.0$\pm$0.5 & 49.0$\pm$1.0 & 52.0$\pm$0.8 \\
\\
1-373161 & 39.86 & 211.0$\pm$0.0 & 139.0$\pm$1.0 & 140.0$\pm$1.0 & 135.0$\pm$1.0 & 124.0$\pm$0.5 & 164.5$\pm$0.5 & 126.0$\pm$0.5 & 33.5$\pm$0.5 \\
1-259650 & 39.47 & 224.0$\pm$1.0 & 234.0$\pm$5.0 & 161.0$\pm$5.0 & 209.0$\pm$50.0 & 166.5$\pm$0.8 & 170.5$\pm$0.5 & 158.5$\pm$0.5 & 166.0$\pm$0.5 \\
1-289865 & 38.73 & 271.0$\pm$2.0 & 128.0$\pm$10.0 & 313.0$\pm$13.0 & 213.0$\pm$10.0 & 40.0$\pm$1.8 & 97.5$\pm$0.5 & 58.5$\pm$0.5 & 22.5$\pm$0.5 \\
\\
1-210646 & 39.84 & 96.0$\pm$1.0 & 108.0$\pm$1.0 & 91.0$\pm$1.0 & 104.0$\pm$1.0 & 106.5$\pm$0.8 & 107.0$\pm$1.0 & 106.5$\pm$0.5 & 106.5$\pm$0.8 \\
1-114306 & 38.65 & 101.0$\pm$1.0 & 44.0$\pm$3.0 & 95.0$\pm$1.0 & 113.0$\pm$3.0 & 128.5$\pm$2.0 & 135.0$\pm$0.8 & 134.0$\pm$0.8 & 133.5$\pm$0.8 \\
1-487130 & 38.93 & 75.0$\pm$1.0 & 50.0$\pm$2.0 & 69.0$\pm$1.0 & 84.0$\pm$2.0 & 97.0$\pm$1.5 & 98.0$\pm$0.5 & 90.5$\pm$0.5 & 95.5$\pm$0.5 \\
\\
\end{tabular}
\end{table*}

\begin{table*}
\contcaption{}
\begin{tabular}{cccccccccc}
\toprule
{\it mangaid} & log$_{10}L_{[{O}{iii}]}$&  $\sigma_{\star}$ &  $\sigma_{[{O}{iii}]}$  &  $\sigma_{H\alpha}$ & $\sigma_{[{N}{ii}]}$ & $\Psi_0$$_{\star}$ & $\Psi_0$ [O\,{\sc iii}]  & $\Psi_0$ H$\alpha$ & $\Psi_0$ [N\,{\sc ii}]  \\
\midrule
1-351790 & 39.83 & 82.0$\pm$1.0 & 88.0$\pm$1.0 & 83.0$\pm$1.0 & 82.0$\pm$2.0 & 135.5$\pm$2.2 & 96.5$\pm$1.8 & 83.5$\pm$2.2 & 91.5$\pm$0.8 \\
1-23731 & 37.83 & 140.0$\pm$1.0 & 106.0$\pm$7.0 & 264.0$\pm$16.0 & 406.0$\pm$54.0 & 134.0$\pm$2.0 & 145.5$\pm$0.8 & 108.0$\pm$0.5 & 126.5$\pm$0.5 \\
1-167334 & 38.83 & 111.0$\pm$1.0 & 89.0$\pm$1.0 & 136.0$\pm$1.0 & 138.0$\pm$1.0 & 120.5$\pm$2.0 & 166.5$\pm$0.5 & 11.0$\pm$34.5 & 56.0$\pm$4.5 \\
\\
1-163831 & 39.64 & 138.0$\pm$1.0 & 173.0$\pm$3.0 & 157.0$\pm$4.0 & 209.0$\pm$3.0 & 95.5$\pm$1.2 & 93.0$\pm$0.5 & 95.5$\pm$1.2 & 97.0$\pm$0.5 \\
1-247456 & 39.3 & 879.0$\pm$34.0 & 172.0$\pm$13.0 & 1090.0$\pm$327.0 & 749.0$\pm$91.0 & 149.0$\pm$1.2 & 169.0$\pm$0.5 & 167.5$\pm$0.8 & 167.5$\pm$0.5 \\
1-210593 & 39.04 & 160.0$\pm$1.0 & 128.0$\pm$3.0 & 146.0$\pm$4.0 & 147.0$\pm$4.0 & 101.5$\pm$1.0 & 107.5$\pm$0.5 & 104.0$\pm$0.8 & 103.0$\pm$0.5 \\
\\
1-22301 & 39.84 & 163.0$\pm$2.0 & 177.0$\pm$4.0 & 137.0$\pm$5.0 & 177.0$\pm$5.0 & 9.5$\pm$1.5 & 8.5$\pm$1.8 & 4.5$\pm$0.5 & 4.0$\pm$0.5 \\
1-251871 & 39.15 & 160.0$\pm$2.0 & 108.0$\pm$5.0 & 128.0$\pm$2.0 & 127.0$\pm$3.0 & 63.0$\pm$0.5 & 66.0$\pm$0.5 & 63.0$\pm$0.5 & 61.5$\pm$1.0 \\
1-72914 & 39.01 & 143.0$\pm$1.0 & 101.0$\pm$4.0 & 110.0$\pm$2.0 & 136.0$\pm$4.0 & 72.5$\pm$0.8 & 82.5$\pm$0.8 & 76.0$\pm$0.5 & 74.0$\pm$0.5 \\
\\
1-248420 & 39.71 & 138.0$\pm$1.0 & 159.0$\pm$1.0 & 152.0$\pm$2.0 & 151.0$\pm$2.0 & 49.5$\pm$1.0 & 49.5$\pm$1.0 & 49.0$\pm$1.0 & 49.0$\pm$1.0 \\
1-211063 & 38.75 & 136.0$\pm$1.0 & 76.0$\pm$2.0 & 141.0$\pm$2.0 & 111.0$\pm$2.0 & 170.0$\pm$1.8 & 164.0$\pm$0.5 & 166.5$\pm$1.8 & 165.0$\pm$1.2 \\
1-211074 & 38.86 & 154.0$\pm$0.0 & 126.0$\pm$2.0 & 150.0$\pm$2.0 & 138.0$\pm$2.0 & 142.5$\pm$1.0 & 133.5$\pm$0.5 & 142.5$\pm$0.8 & 143.0$\pm$0.8 \\
\\
1-23979 & 39.72 & 118.0$\pm$1.0 & 132.0$\pm$1.0 & 115.0$\pm$1.0 & 120.0$\pm$1.0 & 56.5$\pm$1.8 & 67.5$\pm$0.5 & 75.5$\pm$0.5 & 59.0$\pm$0.5 \\
1-320681 & 37.49 & 189.0$\pm$2.0 & 104.0$\pm$22.0 & 288.0$\pm$14.0 & 306.0$\pm$31.0 & 53.0$\pm$0.8 & 67.0$\pm$1.0 & 53.0$\pm$9.8 & 33.0$\pm$0.5 \\
1-519738 & 38.44 & 158.0$\pm$1.0 & 101.0$\pm$3.0 & 135.0$\pm$7.0 & 237.0$\pm$19.0 & 56.5$\pm$1.0 & 52.5$\pm$0.5 & 60.0$\pm$0.5 & 52.5$\pm$0.5 \\
\\
1-542318 & 39.88 & 143.0$\pm$1.0 & 144.0$\pm$2.0 & 172.0$\pm$2.0 & 164.0$\pm$2.0 & 71.0$\pm$2.8 & 119.0$\pm$0.5 & 82.0$\pm$0.5 & 96.0$\pm$0.5 \\
1-285052 & 38.79 & 143.0$\pm$1.0 & 114.0$\pm$4.0 & 137.0$\pm$2.0 & 110.0$\pm$1.0 & 98.0$\pm$2.2 & 75.0$\pm$0.5 & 93.5$\pm$1.2 & 97.0$\pm$1.5 \\
1-377125 & 38.98 & 143.0$\pm$1.0 & 99.0$\pm$2.0 & 144.0$\pm$4.0 & 107.0$\pm$3.0 & 136.5$\pm$9.0 & 100.5$\pm$0.5 & 129.0$\pm$0.8 & 129.0$\pm$0.5 \\
\\
1-95092 & 39.75 & 143.0$\pm$1.0 & 154.0$\pm$3.0 & 130.0$\pm$2.0 & 154.0$\pm$2.0 & 85.5$\pm$1.0 & 80.5$\pm$0.5 & 86.0$\pm$0.5 & 85.5$\pm$0.5 \\
1-210962 & 39.01 & 194.0$\pm$1.0 & 117.0$\pm$2.0 & 147.0$\pm$2.0 & 157.0$\pm$2.0 & 91.5$\pm$1.2 & 96.5$\pm$0.5 & 89.0$\pm$0.5 & 89.5$\pm$0.5 \\
1-251279 & 39.15 & 156.0$\pm$1.0 & 120.0$\pm$2.0 & 117.0$\pm$2.0 & 130.0$\pm$2.0 & 150.0$\pm$1.0 & 142.5$\pm$0.5 & 146.0$\pm$0.5 & 146.0$\pm$0.5 \\
\\
1-279676 & 39.54 & 155.0$\pm$1.0 & 125.0$\pm$2.0 & 120.0$\pm$2.0 & 145.0$\pm$2.0 & 20.0$\pm$1.5 & 27.5$\pm$0.5 & 19.5$\pm$0.5 & 19.5$\pm$0.5 \\
1-44789 & 39.1 & 154.0$\pm$2.0 & 142.0$\pm$3.0 & 173.0$\pm$4.0 & 148.0$\pm$1.0 & 85.0$\pm$0.8 & 95.0$\pm$0.5 & 88.0$\pm$1.5 & 90.0$\pm$0.5 \\
1-378401 & 39.38 & 224.0$\pm$2.0 & 223.0$\pm$4.0 & 188.0$\pm$7.0 & 232.0$\pm$11.0 & 162.5$\pm$1.0 & 33.5$\pm$0.5 & 38.0$\pm$0.5 & 26.5$\pm$0.5 \\
\\
1-201561 & 39.59 & 168.0$\pm$2.0 & 157.0$\pm$2.0 & 168.0$\pm$2.0 & 166.0$\pm$1.0 & 23.5$\pm$3.0 & 22.5$\pm$0.5 & 29.5$\pm$0.5 & 42.5$\pm$0.5 \\
1-24246 & 39.1 & 110.0$\pm$1.0 & 134.0$\pm$4.0 & 94.0$\pm$2.0 & 105.0$\pm$2.0 & 139.0$\pm$1.5 & 134.0$\pm$0.8 & 134.0$\pm$1.0 & 121.5$\pm$0.5 \\
1-285052 & 38.79 & 143.0$\pm$1.0 & 114.0$\pm$4.0 & 137.0$\pm$2.0 & 110.0$\pm$1.0 & 98.0$\pm$2.2 & 75.0$\pm$0.5 & 93.5$\pm$1.2 & 97.0$\pm$1.5 \\
\\
1-198182 & 39.68 & 217.0$\pm$1.0 & 195.0$\pm$1.0 & 219.0$\pm$1.0 & 227.0$\pm$2.0 & 44.5$\pm$0.8 & 35.5$\pm$0.8 & 42.5$\pm$0.8 & 55.0$\pm$0.5 \\
1-256185 & 39.22 & 214.0$\pm$1.0 & 171.0$\pm$10.0 & 151.0$\pm$5.0 & 202.0$\pm$12.0 & 178.5$\pm$0.8 & 162.5$\pm$0.8 & 167.5$\pm$0.8 & 168.0$\pm$1.2 \\
1-48053 & -- & 286.0$\pm$1.0 & 116.0$\pm$36.0 & 184.0$\pm$10.0 & 294.0$\pm$13.0 & 12.0$\pm$0.8 & 161.5$\pm$0.5 & 39.0$\pm$0.5 & 160.0$\pm$0.5 \\
\\
1-96075 & 39.53 & 122.0$\pm$1.0 & 133.0$\pm$2.0 & 115.0$\pm$2.0 & 137.0$\pm$2.0 & 42.0$\pm$0.8 & 47.5$\pm$0.5 & 43.5$\pm$0.5 & 44.0$\pm$0.5 \\
1-166947 & 38.79 & 149.0$\pm$1.0 & 83.0$\pm$3.0 & 143.0$\pm$3.0 & 116.0$\pm$2.0 & 36.0$\pm$3.2 & 123.0$\pm$0.5 & 34.0$\pm$0.8 & 35.5$\pm$0.8 \\
1-52259 & 38.98 & 89.0$\pm$1.0 & 100.0$\pm$3.0 & 63.0$\pm$0.0 & 63.0$\pm$0.0 & 90.5$\pm$4.2 & 103.0$\pm$0.5 & 87.5$\pm$12.0 & 99.5$\pm$6.2 \\
\\
1-519742 & 39.78 & 96.0$\pm$1.0 & 99.0$\pm$0.0 & 86.0$\pm$1.0 & 80.0$\pm$1.0 & 133.0$\pm$4.0 & 124.5$\pm$0.5 & 129.5$\pm$1.8 & 120.5$\pm$1.0 \\
1-37079 & 37.71 & 72.0$\pm$1.0 & 39.0$\pm$3.0 & 78.0$\pm$1.0 & 96.0$\pm$2.0 & 157.5$\pm$6.0 & 15.0$\pm$0.5 & 177.5$\pm$3.0 & 25.5$\pm$0.5 \\
1-276679 & 38.42 & 88.0$\pm$1.0 & 99.0$\pm$4.0 & 61.0$\pm$0.0 & 71.0$\pm$0.0 & 152.0$\pm$2.2 & 137.5$\pm$0.5 & 151.5$\pm$0.8 & 152.5$\pm$0.5 \\
\\
1-491229 & 39.64 & 189.0$\pm$1.0 & 213.0$\pm$5.0 & 249.0$\pm$3.0 & 226.0$\pm$2.0 & 101.5$\pm$1.2 & 82.5$\pm$0.5 & 99.5$\pm$0.5 & 105.0$\pm$0.5 \\
1-94554 & 38.98 & 136.0$\pm$1.0 & 111.0$\pm$3.0 & 135.0$\pm$2.0 & 146.0$\pm$5.0 & 29.5$\pm$2.0 & 37.5$\pm$0.5 & 8.0$\pm$0.8 & 44.0$\pm$0.5 \\
1-604048 & 39.14 & 163.0$\pm$1.0 & 126.0$\pm$1.0 & 124.0$\pm$2.0 & 129.0$\pm$3.0 & 58.5$\pm$2.5 & 71.0$\pm$0.5 & 59.0$\pm$1.8 & 67.0$\pm$1.8 \\
\\
1-604761 & 39.64 & 193.0$\pm$2.0 & 186.0$\pm$3.0 & 178.0$\pm$2.0 & 189.0$\pm$3.0 & 70.0$\pm$0.5 & 68.0$\pm$0.5 & 66.5$\pm$0.5 & 66.5$\pm$0.5 \\
1-210173 & 39.28 & 155.0$\pm$2.0 & 137.0$\pm$4.0 & 113.0$\pm$2.0 & 135.0$\pm$3.0 & 122.5$\pm$1.0 & 121.0$\pm$0.5 & 121.0$\pm$0.5 & 124.5$\pm$0.5 \\
1-71525 & 39.0 & 136.0$\pm$1.0 & 94.0$\pm$5.0 & 242.0$\pm$135.0 & 685.0$\pm$263.0 & 126.5$\pm$1.2 & 138.0$\pm$0.5 & 130.5$\pm$0.8 & 130.5$\pm$0.5 \\
\\
1-25725 & 39.39 & 190.0$\pm$2.0 & 198.0$\pm$3.0 & 195.0$\pm$4.0 & 225.0$\pm$4.0 & 138.5$\pm$2.2 & 136.5$\pm$0.5 & 133.0$\pm$1.2 & 144.5$\pm$0.5 \\
1-211079 & 38.53 & 198.0$\pm$1.0 & 228.0$\pm$11.0 & 250.0$\pm$18.0 & 252.0$\pm$15.0 & 178.5$\pm$0.8 & 167.0$\pm$0.5 & 126.5$\pm$0.5 & 125.5$\pm$0.5 \\
1-322074 & 38.12 & 120.0$\pm$0.0 & 93.0$\pm$3.0 & 227.0$\pm$13.0 & 234.0$\pm$49.0 & 35.5$\pm$2.5 & 68.5$\pm$0.5 & 120.0$\pm$0.5 & 160.5$\pm$0.5 \\
\\
1-94604 & 39.5 & 110.0$\pm$1.0 & 109.0$\pm$1.0 & 136.0$\pm$4.0 & 343.0$\pm$54.0 & 153.0$\pm$2.0 & 166.0$\pm$0.5 & 157.5$\pm$0.5 & 144.0$\pm$0.5 \\
1-295095 & 38.76 & 101.0$\pm$1.0 & 112.0$\pm$3.0 & 80.0$\pm$1.0 & 88.0$\pm$2.0 & 1.0$\pm$2.2 & 3.0$\pm$0.8 & 178.0$\pm$1.5 & 1.0$\pm$1.2 \\
1-134239 & 39.17 & 138.0$\pm$1.0 & 136.0$\pm$3.0 & 119.0$\pm$3.0 & 140.0$\pm$3.0 & 32.5$\pm$1.5 & 31.5$\pm$0.5 & 34.0$\pm$1.5 & 35.5$\pm$1.5 \\
\\
\end{tabular}
\end{table*}

\begin{table*}
\contcaption{}
\begin{tabular}{cccccccccc}
\toprule
{\it mangaid} & log$_{10}L_{[{O}{iii}]}$&  $\sigma_{\star}$ &  $\sigma_{[{O}{iii}]}$  &  $\sigma_{H\alpha}$ & $\sigma_{[{N}{ii}]}$ & $\Psi_0$$_{\star}$ & $\Psi_0$ [O\,{\sc iii}]  & $\Psi_0$ H$\alpha$ & $\Psi_0$ [N\,{\sc ii}]  \\
\midrule
1-37036 & 39.42 & 220.0$\pm$1.0 & 173.0$\pm$2.0 & 206.0$\pm$2.0 & 197.0$\pm$3.0 & 98.0$\pm$0.8 & 86.0$\pm$0.5 & 84.0$\pm$0.5 & 99.0$\pm$0.5 \\
1-210785 & 37.08 & 165.0$\pm$1.0 & 102.0$\pm$12.0 & 569.0$\pm$171.0 & 2128.0$\pm$444.0 & 125.5$\pm$0.8 & 130.0$\pm$0.5 & 113.5$\pm$0.5 & 107.0$\pm$0.5 \\
1-25680 & 39.08 & 230.0$\pm$1.0 & 140.0$\pm$3.0 & 176.0$\pm$3.0 & 189.0$\pm$4.0 & 55.5$\pm$2.0 & 168.0$\pm$0.5 & 1.0$\pm$0.5 & 174.5$\pm$0.5 \\
\\
1-167688 & 39.56 & 96.0$\pm$1.0 & 81.0$\pm$0.0 & 95.0$\pm$0.0 & 89.0$\pm$1.0 & 35.5$\pm$9.2 & 18.0$\pm$1.0 & 40.0$\pm$0.8 & 70.5$\pm$1.5 \\
1-235587 & 38.59 & 125.0$\pm$0.0 & 84.0$\pm$1.0 & 105.0$\pm$2.0 & 106.0$\pm$2.0 & 85.5$\pm$2.2 & 92.0$\pm$0.5 & 99.0$\pm$1.8 & 100.0$\pm$1.0 \\
1-37062 & 39.02 & 79.0$\pm$1.0 & 83.0$\pm$1.0 & 91.0$\pm$1.0 & 96.0$\pm$1.0 & 21.5$\pm$3.5 & 173.0$\pm$0.5 & 19.0$\pm$12.0 & 31.0$\pm$0.5 \\
\\
1-279666 & 39.45 & 115.0$\pm$1.0 & 103.0$\pm$1.0 & 104.0$\pm$1.0 & 121.0$\pm$1.0 & 57.5$\pm$2.2 & 52.0$\pm$0.5 & 60.0$\pm$1.8 & 46.5$\pm$1.5 \\
1-392976 & 38.57 & 108.0$\pm$1.0 & 94.0$\pm$3.0 & 133.0$\pm$5.0 & 123.0$\pm$4.0 & 16.0$\pm$0.5 & 56.0$\pm$0.5 & 80.5$\pm$0.5 & 166.0$\pm$0.5 \\
1-47499 & 38.86 & 110.0$\pm$0.0 & 102.0$\pm$3.0 & 128.0$\pm$4.0 & 121.0$\pm$4.0 & 108.0$\pm$2.2 & 82.5$\pm$1.8 & 108.0$\pm$0.5 & 126.0$\pm$0.5 \\
\\
1-339163 & 39.35 & 159.0$\pm$1.0 & 323.0$\pm$6.0 & 245.0$\pm$4.0 & 353.0$\pm$8.0 & 38.5$\pm$0.5 & 42.0$\pm$0.5 & 39.0$\pm$0.5 & 39.0$\pm$0.5 \\
1-136125 & 38.56 & 109.0$\pm$1.0 & 61.0$\pm$2.0 & 118.0$\pm$2.0 & 112.0$\pm$2.0 & 68.0$\pm$0.5 & 69.0$\pm$0.5 & 67.5$\pm$0.5 & 67.5$\pm$0.5 \\
1-626830 & 38.62 & 137.0$\pm$1.0 & 107.0$\pm$3.0 & 115.0$\pm$2.0 & 136.0$\pm$2.0 & 22.5$\pm$0.5 & 29.5$\pm$0.5 & 22.5$\pm$0.5 & 19.5$\pm$0.5 \\
\\
1-258774 & 39.4 & 143.0$\pm$0.0 & 114.0$\pm$2.0 & 110.0$\pm$1.0 & 135.0$\pm$0.0 & 154.0$\pm$4.2 & 152.0$\pm$1.2 & 155.0$\pm$3.2 & 172.5$\pm$3.8 \\
1-379660 & 39.12 & 139.0$\pm$1.0 & 132.0$\pm$3.0 & 118.0$\pm$4.0 & 127.0$\pm$4.0 & 115.5$\pm$1.2 & 106.0$\pm$0.5 & 111.5$\pm$0.8 & 113.5$\pm$0.5 \\
1-48208 & 38.7 & 162.0$\pm$1.0 & 94.0$\pm$2.0 & 140.0$\pm$4.0 & 91.0$\pm$5.0 & 71.5$\pm$2.2 & 75.0$\pm$0.5 & 76.0$\pm$0.5 & 66.5$\pm$0.5 \\
\\
1-198153 & 39.29 & 176.0$\pm$1.0 & 145.0$\pm$2.0 & 177.0$\pm$1.0 & 178.0$\pm$1.0 & 93.5$\pm$0.5 & 96.0$\pm$0.5 & 94.5$\pm$0.5 & 94.5$\pm$0.5 \\
1-211063 & 38.75 & 136.0$\pm$1.0 & 76.0$\pm$2.0 & 141.0$\pm$2.0 & 111.0$\pm$2.0 & 170.0$\pm$1.8 & 164.0$\pm$0.5 & 166.5$\pm$1.8 & 165.0$\pm$1.2 \\
1-135810 & 38.45 & 112.0$\pm$1.0 & 80.0$\pm$2.0 & 95.0$\pm$2.0 & 101.0$\pm$2.0 & 52.0$\pm$0.8 & 54.5$\pm$0.5 & 50.5$\pm$0.8 & 50.5$\pm$0.5 \\
\\
1-91016 & 39.49 & 103.0$\pm$1.0 & 125.0$\pm$2.0 & 138.0$\pm$3.0 & 173.0$\pm$4.0 & 50.5$\pm$4.0 & 44.5$\pm$1.2 & 51.5$\pm$1.2 & 57.0$\pm$2.0 \\
1-338828 & 39.3 & 94.0$\pm$1.0 & 90.0$\pm$1.0 & 80.0$\pm$1.0 & 80.0$\pm$1.0 & 103.0$\pm$0.5 & 101.5$\pm$0.5 & 96.0$\pm$0.5 & 86.0$\pm$1.2 \\
1-386695 & 39.68 & 105.0$\pm$2.0 & 119.0$\pm$1.0 & 105.0$\pm$1.0 & 107.0$\pm$1.0 & 129.0$\pm$1.2 & 88.0$\pm$0.5 & 141.5$\pm$0.5 & 89.5$\pm$0.5 \\
\\
1-279073 & 39.34 & 231.0$\pm$1.0 & 161.0$\pm$4.0 & 186.0$\pm$2.0 & 181.0$\pm$3.0 & 15.0$\pm$1.5 & 169.5$\pm$0.8 & 155.5$\pm$0.5 & 164.0$\pm$0.5 \\
1-211100 & 38.09 & 198.0$\pm$1.0 & 116.0$\pm$7.0 & 275.0$\pm$44.0 & 346.0$\pm$14.0 & 152.0$\pm$4.8 & 127.0$\pm$0.5 & 103.0$\pm$0.5 & 165.0$\pm$0.5 \\
1-210784 & 38.3 & 171.0$\pm$0.0 & 97.0$\pm$4.0 & 159.0$\pm$2.0 & 218.0$\pm$3.0 & 90.0$\pm$2.2 & 101.0$\pm$0.5 & 59.0$\pm$1.2 & 62.5$\pm$0.5 \\
\\
1-135044 & 39.44 & 115.0$\pm$1.0 & 116.0$\pm$2.0 & 156.0$\pm$2.0 & 167.0$\pm$3.0 & 95.5$\pm$1.2 & 96.5$\pm$0.5 & 96.5$\pm$1.2 & 98.5$\pm$0.5 \\
1-218280 & 38.7 & 148.0$\pm$0.0 & 104.0$\pm$1.0 & 121.0$\pm$1.0 & 114.0$\pm$3.0 & 99.0$\pm$1.0 & 96.0$\pm$0.5 & 98.0$\pm$0.5 & 98.0$\pm$0.5 \\
1-211063 & 38.75 & 136.0$\pm$1.0 & 76.0$\pm$2.0 & 141.0$\pm$2.0 & 111.0$\pm$2.0 & 170.0$\pm$1.8 & 164.0$\pm$0.5 & 166.5$\pm$1.8 & 165.0$\pm$1.2 \\
\\
1-148068 & 39.21 & 156.0$\pm$1.0 & 121.0$\pm$4.0 & 169.0$\pm$2.0 & 156.0$\pm$1.0 & 63.5$\pm$0.8 & 59.0$\pm$0.5 & 65.0$\pm$0.8 & 66.5$\pm$0.5 \\
1-166947 & 38.79 & 149.0$\pm$1.0 & 83.0$\pm$3.0 & 143.0$\pm$3.0 & 116.0$\pm$2.0 & 36.0$\pm$3.2 & 123.0$\pm$0.5 & 34.0$\pm$0.8 & 35.5$\pm$0.8 \\
1-55572 & 38.93 & 194.0$\pm$2.0 & 117.0$\pm$1.0 & 208.0$\pm$4.0 & 198.0$\pm$5.0 & 51.0$\pm$0.5 & 52.0$\pm$0.5 & 50.5$\pm$0.5 & 52.0$\pm$0.8 \\
\\
1-277552 & 39.27 & 118.0$\pm$1.0 & 119.0$\pm$1.0 & 99.0$\pm$1.0 & 112.0$\pm$1.0 & 3.5$\pm$0.5 & 4.0$\pm$0.5 & 2.5$\pm$0.5 & 2.5$\pm$0.8 \\
1-264513 & 39.28 & 99.0$\pm$1.0 & 106.0$\pm$1.0 & 77.0$\pm$0.0 & 85.0$\pm$1.0 & 0.5$\pm$0.5 & 177.0$\pm$3.8 & 175.5$\pm$1.5 & 175.5$\pm$2.0 \\
1-136125 & 38.56 & 109.0$\pm$1.0 & 61.0$\pm$2.0 & 118.0$\pm$2.0 & 112.0$\pm$2.0 & 68.0$\pm$0.5 & 69.0$\pm$0.5 & 67.5$\pm$0.5 & 67.5$\pm$0.5 \\
\\
1-217050 & 39.05 & 203.0$\pm$1.0 & 154.0$\pm$2.0 & 200.0$\pm$2.0 & 208.0$\pm$3.0 & 60.5$\pm$0.5 & 83.0$\pm$0.5 & 82.0$\pm$0.5 & 78.0$\pm$0.5 \\
1-135372 & 38.66 & 212.0$\pm$1.0 & 183.0$\pm$6.0 & 1011.0$\pm$363.0 & 280.0$\pm$13.0 & 20.0$\pm$0.8 & 23.0$\pm$0.5 & 22.5$\pm$0.5 & 29.5$\pm$0.5 \\
1-274663 & 38.87 & 216.0$\pm$1.0 & 125.0$\pm$3.0 & 188.0$\pm$3.0 & 188.0$\pm$5.0 & 153.5$\pm$1.8 & 163.5$\pm$0.5 & 111.5$\pm$0.5 & 155.5$\pm$0.5 \\
\\
1-25554 & 38.96 & 114.0$\pm$0.0 & 111.0$\pm$1.0 & 117.0$\pm$1.0 & 112.0$\pm$2.0 & 74.5$\pm$1.2 & 67.5$\pm$0.5 & 76.5$\pm$0.5 & 82.0$\pm$0.5 \\
1-135625 & 39.33 & 98.0$\pm$1.0 & 111.0$\pm$2.0 & 98.0$\pm$1.0 & 98.0$\pm$1.0 & 17.5$\pm$1.2 & 16.0$\pm$1.5 & 15.0$\pm$0.5 & 12.5$\pm$1.0 \\
1-216958 & 38.9 & 99.0$\pm$0.0 & 78.0$\pm$0.0 & 67.0$\pm$0.0 & 72.0$\pm$0.0 & 58.0$\pm$5.8 & 41.0$\pm$0.5 & 60.0$\pm$4.5 & 81.5$\pm$0.5 \\
\\
1-135285 & 38.93 & 133.0$\pm$1.0 & 155.0$\pm$2.0 & 166.0$\pm$2.0 & 162.0$\pm$1.0 & 116.5$\pm$0.5 & 112.0$\pm$0.5 & 114.0$\pm$0.5 & 113.0$\pm$0.5 \\
1-633990 & 38.77 & 97.0$\pm$0.0 & 93.0$\pm$1.0 & 99.0$\pm$0.0 & 103.0$\pm$1.0 & 33.5$\pm$1.2 & 30.0$\pm$0.5 & 31.5$\pm$0.5 & 37.5$\pm$0.5 \\
1-25688 & 38.35 & 110.0$\pm$1.0 & 64.0$\pm$2.0 & 110.0$\pm$2.0 & 115.0$\pm$1.0 & 20.5$\pm$1.5 & 13.5$\pm$0.5 & 17.0$\pm$0.8 & 19.5$\pm$1.5 \\
\bottomrule
\end{tabular}
\end{table*}

\newpage
\begin{table*}
\caption{Parameters of AGN in MaNGA-MPL5. (1) galaxy identification in the MaNGA survey;  (2)-(3): RA/DEC (2000) in degrees; (4) spectroscopic redshift from SDSS-III; (5): integrated
absolute $r$-band magnitude from SDSS-III; (6): stellar mass in units of $M_\odot$; errors associated to the stellar masses of galaxies in our sample are typically under 0.03\,dex \citep{Conroy+09}; (7) elliptical/spiral/merging classification from Galaxy Zoo I; (8)-(9): $r$-band concentration and asymmetry; (10) [OIII] luminosity in units of $10^{40}$\,erg\,s$^{-1}$. Table extracted of \citet{rembold17} .}
\begin{tabular}{ccccccccccccc}
\hline
mangaID & RA  & DEC & $z$ & $M_r$ & log $M^\star/M_\odot$ & GZ1$_c$ & $C$ & $A$  & $L(\rm{[OIII]})$\\
 (1)    & (2)      & (3) & (4) & (5) & (6)   & (7)                   & (8)     & (9) & (10) \\
\hline
1-558912 & 166.129410 & 42.624554 & 0.1261 & -20.46 & 11.25 & -- & 0.37 & 0.12 & 56.82$\pm$1.25\\
1-269632 & 247.560974 & 26.206474 & 0.1315 & -21.78 & 11.62 & S & 0.47 & 0.05 & 30.08$\pm$1.69\\
1-258599 & 186.181000 & 44.410770 & 0.1256 & -21.24 & 11.68 & E & 0.50 & 0.11 & 20.95$\pm$0.67\\
1-72322 & 121.014198 & 40.802612 & 0.1262 & -21.81 & 12.05 & S & 0.34 & 0.08 & 20.66$\pm$0.43\\
1-121532 & 118.091110 & 34.326569 & 0.1400 & -20.51 & 11.34 & E & 0.33 & 0.05 & 11.68$\pm$0.96\smallskip\\
1-209980 & 240.470871 & 45.351940 & 0.0420 & -19.70 & 10.79 & S & 0.57 & 0.04 & 11.01$\pm$0.17\\
1-44379 & 120.700706 & 45.034554 & 0.0389 & -19.89 & 10.97 & S & 0.24 & 0.06 & 8.94$\pm$0.14\\
1-149211 & 168.947800 & 50.401634 & 0.0473 & -18.27 & 10.16 & S & 0.29 & 0.03 & 7.88$\pm$0.14\\
1-173958 & 167.306015 & 49.519432 & 0.0724 & -20.53 & 11.31 & S & 0.33 & 0.06 & 6.79$\pm$0.30\\
1-338922 & 114.775749 & 44.402767 & 0.1345 & -20.27 & 11.13 & M & 0.44 & 0.03 & 6.77$\pm$0.90\smallskip\\
1-279147 & 168.957733 & 46.319565 & 0.0533 & -19.51 & 10.66 & S & 0.45 & 0.03 & 6.77$\pm$0.20\\
1-460812 & 127.170799 & 17.581400 & 0.0665 & -19.81 & 11.44 & -- & 0.38 & 0.05 & 6.46$\pm$0.31\\
1-92866 & 243.581818 & 50.465611 & 0.0603 & -20.56 & 11.69 & E & 0.49 & 0.05 & 6.12$\pm$0.30\\
1-94784 & 249.318420 & 44.418228 & 0.0314 & -20.06 & 10.85 & S & 0.42 & 0.03 & 5.96$\pm$0.12\\
1-44303 & 119.182152 & 44.856709 & 0.0499 & -19.72 & 10.62 & S & 0.29 & 0.10 & 5.56$\pm$0.12\smallskip\\
1-339094 & 117.472420 & 45.248482 & 0.0313 & -19.02 & 10.52 & E & 0.36 & 0.03 & 5.29$\pm$0.09\\
1-137883 & 137.874756 & 45.468319 & 0.0268 & -18.06 & 10.77 & E/S & 0.41 & 0.01 & 3.87$\pm$0.12\\
1-48116 & 132.653992 & 57.359669 & 0.0261 & -19.18 & 10.60 & S & 0.31 & 0.06 & 3.79$\pm$0.08\\
1-256446 & 166.509872 & 43.173473 & 0.0584 & -19.40 & 11.14 & E & 0.49 & 0.05 & 3.74$\pm$0.15\\
1-95585 & 255.029877 & 37.839500 & 0.0633 & -20.88 & 11.24 & S & 0.27 & 0.08 & 3.58$\pm$0.16\smallskip\\
1-135641 & 249.557312 & 40.146820 & 0.0304 & -19.03 & 11.19 & S & 0.28 & 0.08 & 3.52$\pm$0.09\\
1-259142 & 193.703995 & 44.155567 & 0.0543 & -20.75 & 11.29 & S & 0.39 & 0.06 & 3.47$\pm$0.20\\
1-109056 & 39.446587 & 0.405085 & 0.0473 & -19.27 & 10.57 & -- & 0.32 & 0.05 & 3.24$\pm$0.08\\
1-24148 & 258.827423 & 57.658772 & 0.0282 & -18.51 & 10.56 & S & 0.31 & 0.04 & 3.17$\pm$0.05\\
1-166919 & 146.709106 & 43.423843 & 0.0722 & -20.85 & 11.28 & S & 0.37 & 0.06 & 2.64$\pm$0.25\smallskip\\
1-248389 & 240.658051 & 41.293427 & 0.0348 & -19.36 & 10.57 & S & 0.49 & 0.12 & 2.55$\pm$0.09\\
1-321739 & 226.431656 & 44.404903 & 0.0283 & -18.91 & 11.12 & S & 0.40 & 0.14 & 2.24$\pm$0.10\\
1-234618 & 202.128433 & 47.714039 & 0.0608 & -19.64 & 11.37 & S & 0.31 & 0.09 & 2.23$\pm$0.23\\
1-229010 & 57.243038 & -1.144831 & 0.0407 & -20.51 & 11.46 & -- & 0.41 & 0.03 & 2.11$\pm$0.09\\
1-211311 & 248.426392 & 39.185120 & 0.0298 & -19.04 & 10.44 & E/S & 0.43 & 0.02 & 1.99$\pm$0.06\smallskip\\
1-373161 & 222.810074 & 30.692245 & 0.0547 & -21.30 & 11.60 & E & 0.43 & 0.00 & 1.87$\pm$0.11\\
1-210646 & 245.157181 & 41.466873 & 0.0606 & -20.38 & 10.98 & S & 0.18 & 0.05 & 1.80$\pm$0.10\\
1-351790 & 121.147926 & 50.708557 & 0.0227 & -18.09 & 9.92 & E & 0.39 & 0.02 & 1.72$\pm$0.03\\
1-163831 & 118.627846 & 25.815987 & 0.0631 & -20.84 & 11.26 & S & 0.27 & 0.05 & 1.67$\pm$0.13\\
1-22301 & 253.405563 & 63.031269 & 0.1052 & -21.19 & 11.18 & S & 0.29 & 0.08 & 1.67$\pm$0.23\smallskip\\
1-248420 & 241.823395 & 41.403603 & 0.0346 & -19.71 & 10.90 & S & 0.21 & 0.07 & 1.66$\pm$0.06\\
1-23979 & 258.158752 & 57.322422 & 0.0266 & -18.27 & 10.42 & E & 0.44 & 0.06 & 1.60$\pm$0.05\\
1-542318 & 245.248306 & 49.001778 & 0.0582 & -19.75 & 10.91 & E & 0.34 & 0.01 & 1.58$\pm$0.07\\
1-95092 & 250.846420 & 39.806461 & 0.0302 & -19.95 & 11.20 & E & 0.47 & 0.04 & 1.54$\pm$0.07\\
1-279676 & 173.981888 & 48.021458 & 0.0587 & -19.40 & 10.81 & -- & 0.32 & 0.02 & 1.52$\pm$0.14\smallskip\\
1-201561 & 118.053215 & 28.772579 & 0.0637 & -19.73 & 10.88 & S & 0.30 & 0.07 & 1.37$\pm$0.15\\
1-198182 & 224.749649 & 48.409855 & 0.0359 & -20.22 & 11.09 & E & 0.49 & 0.01 & 1.34$\pm$0.11\\
1-96075 & 253.946381 & 39.310535 & 0.0631 & -21.12 & 11.35 & S & 0.29 & 0.07 & 1.26$\pm$0.13\\
1-519742 & 206.612457 & 22.076742 & 0.0276 & -17.62 & 9.64 & S & 0.23 & 0.04 & 1.19$\pm$0.03\\
1-491229 & 172.607544 & 22.216530 & 0.0393 & -20.25 & 11.12 & E & 0.51 & 0.02 & 1.14$\pm$0.11\smallskip\\
1-604761 & 113.472275 & 37.025906 & 0.0618 & -20.92 & 11.34 & S & 0.26 & 0.12 & 1.00$\pm$0.13\\
1-25725 & 262.996735 & 59.971638 & 0.0291 & -18.30 & 10.55 & E & 0.44 & 0.04 & 0.92$\pm$0.05\\
1-94604 & 251.335938 & 42.757790 & 0.0493 & -19.44 & 10.52 & S & 0.37 & 0.01 & 0.86$\pm$0.07\\
1-37036 & 41.699909 & 0.421577 & 0.0283 & -19.02 & 10.66 & E & 0.40 & 0.09 & 0.84$\pm$0.06\\
1-167688 & 155.885559 & 46.057755 & 0.0258 & -17.86 & 9.75 & E & 0.52 & 0.04 & 0.84$\pm$0.02\smallskip\\
1-279666 & 173.911240 & 47.515518 & 0.0455 & -18.83 & 10.42 & E & 0.31 & 0.02 & 0.84$\pm$0.07\\
1-339163 & 116.280205 & 46.072422 & 0.0312 & -20.02 & 10.97 & S & 0.30 & 0.10 & 0.82$\pm$0.07\\
1-258774 & 186.400864 & 45.083858 & 0.0384 & -19.60 & 10.77 & -- & 0.55 & 0.03 & 0.77$\pm$0.10\\
1-198153 & 224.289078 & 48.633968 & 0.0354 & -19.83 & 11.00 & S & 0.27 & 0.07 & 0.76$\pm$0.08\\
1-91016 & 234.810974 & 56.670856 & 0.0463 & -18.60 & 10.56 & S & 0.27 & 0.06 & 0.76$\pm$0.09\smallskip\\
1-279073 & 170.588150 & 46.430504 & 0.0323 & -19.53 & 10.79 & E & 0.51 & 0.01 & 0.63$\pm$0.06\\
1-135044 & 247.907990 & 41.493645 & 0.0303 & -19.76 & 10.65 & S & 0.31 & 0.05 & 0.61$\pm$0.04\\
\hline
\end{tabular}
\label{tableagns}
\end{table*}

\begin{table*}
\contcaption{}
\begin{tabular}{ccccccccccccc}
\hline
mangaID & RA  & DEC & $z$ & $M_r$ & log $M^\star/M_\odot$ & GZ1$_c$ & $C$ & $A$  & $L(\rm{[OIII]})$\\
 (1)    & (2)      & (3) & (4) & (5) & (6)   & (7)                   & (8)     & (9) & (10) \\
\hline
1-148068 & 156.805679 & 48.244793 & 0.0610 & -20.72 & 11.41 & S & 0.22 & 0.04 & 0.45$\pm$0.15\\
1-277552 & 167.034561 & 45.984623 & 0.0362 & -19.72 & 10.83 & S & 0.21 & 0.15 & 0.44$\pm$0.05\\
1-217050 & 136.719986 & 41.408253 & 0.0274 & -19.66 & 10.93 & E & 0.47 & 0.02 & 0.43$\pm$0.03\smallskip\\
1-25554 & 262.486053 & 58.397408 & 0.0268 & -19.27 & 10.52 & S & 0.36 & 0.04 & 0.24$\pm$0.03\\
1-135285 & 247.216949 & 42.812012 & 0.0316 & -19.66 & 10.78 & -- & 0.32 & 0.05 & 0.20$\pm$0.04\\
\hline
\end{tabular}
\end{table*}

\newpage
\begin{table*}
\caption{Control sample parameters. (1) identification of the AGN host associated to the control galaxy; (2)-(11) same as (1)-(10) of Table~\ref{tableagns}. Twelve control sample objects
have been paired to two different AGN hosts and appear more than once in the table. Table extracted of \citet{rembold17} .}
\begin{tabular}{ccccccccccccc}
\hline
AGN mangaID & mangaID & RA  & DEC & $z$ & $M_r$ & log $M^\star/M_\odot$ & GZ1$_c$ & $C$ & $A$  & $L(\rm{[OIII]})$\\
 (1)    & (2)      & (3) & (4) & (5) & (6)   & (7)                   & (8)     & (9) & (10) & (11) \\
\hline
1-558912 & 1-71481 & 117.456001 & 34.883911 & 0.1312 & -20.95 & 11.70 & E & 0.47 & 0.02 & 0.10$\pm$0.20\\
         & 1-72928 & 127.256485 & 45.016773 & 0.1270 & -20.62 & 11.52 & E & 0.40 & 0.21 & 0.09$\pm$0.23\\
1-269632 & 1-210700 & 248.140564 & 39.131020 & 0.1303 & -20.96 & 11.67 & S & 0.36 & 0.03 & 1.55$\pm$0.44\\
         & 1-378795 & 118.925613 & 50.172771 & 0.0967 & -20.77 & 11.35 & S & 0.32 & 0.03 & 0.72$\pm$0.31\smallskip\\
1-258599 & 1-93876 & 246.942947 & 44.177521 & 0.1394 & -20.75 & 11.50 & E & 0.44 & 0.01 & 0.46$\pm$0.36\\
         & 1-166691 & 146.047348 & 42.900040 & 0.1052 & -20.50 & 11.36 & E & 0.51 & 0.04 & 0.09$\pm$0.49\\
1-72322 & 1-121717 & 118.803429 & 35.596798 & 0.1098 & -21.11 & 11.61 & S & 0.39 & 0.12 & 1.40$\pm$0.57\\
         & 1-43721 & 116.967567 & 43.383499 & 0.1114 & -21.41 & 11.86 & S & 0.32 & 0.01 & 1.91$\pm$0.52\smallskip\\
1-121532 & 1-218427 & 124.342316 & 27.796206 & 0.1496 & -21.30 & 11.47 & E & 0.47 & 0.04 & 0.72$\pm$0.62\\
         & 1-177493 & 257.085754 & 31.746916 & 0.1081 & -20.90 & 11.30 & E & 0.38 & 0.06 & 2.29$\pm$0.28\\
1-209980 & 1-295095 & 248.348663 & 24.776577 & 0.0410 & -18.40 & 10.14 & E & 0.35 & 0.05 & 0.15$\pm$0.03\\
         & 1-92626 & 241.799545 & 48.572563 & 0.0434 & -20.04 & 11.04 & S & 0.36 & 0.03 & 0.76$\pm$0.07\smallskip\\
1-44379 & 1-211082 & 247.620041 & 39.626045 & 0.0304 & -19.72 & 11.07 & E & 0.31 & 0.06 & 0.19$\pm$0.04\\
         & 1-135371 & 250.156235 & 39.221634 & 0.0352 & -19.20 & 10.76 & S & 0.28 & 0.11 & 0.25$\pm$0.07\\
1-149211 & 1-377321 & 110.556152 & 42.183643 & 0.0444 & -19.02 & 9.89 & S & 0.31 & 0.03 & 4.53$\pm$0.13\\
         & 1-491233 & 172.563995 & 22.992010 & 0.0332 & -18.39 & 10.59 & S & 0.29 & 0.06 & 0.25$\pm$0.03\smallskip\\
1-173958 & 1-247456 & 232.823196 & 45.416538 & 0.0705 & -20.05 & 10.83 & -- & 0.40 & 0.02 & 0.57$\pm$0.16\\
         & 1-24246 & 264.840790 & 56.567070 & 0.0818 & -19.91 & 10.57 & S & 0.75 & 0.36 & 0.11$\pm$0.06\\
1-338922 & 1-286804 & 211.904861 & 44.482269 & 0.1429 & -20.03 & 10.50 & M & 0.44 & 0.32 & 2.23$\pm$0.43\\
         & 1-109493 & 56.425140 & -0.378460 & 0.1093 & -20.46 & 11.26 & -- & 0.49 & -0.01 & 0.15$\pm$0.18\smallskip\\
1-279147 & 1-283246 & 191.078873 & 46.407131 & 0.0496 & -19.17 & 10.55 & S & 0.47 & 0.04 & 0.23$\pm$0.06\\
         & 1-351538 & 119.145126 & 47.563850 & 0.0692 & -19.67 & 11.00 & S & 0.35 & 0.08 & 0.46$\pm$0.13\\
1-460812 & 1-270160 & 248.274612 & 26.211815 & 0.0660 & -20.37 & 11.46 & S & 0.50 & 0.02 & 0.70$\pm$0.39\\
         & 1-258455 & 183.612198 & 45.195454 & 0.0653 & -20.02 & 11.03 & E & 0.40 & 0.03 & 0.49$\pm$0.14\smallskip\\
1-92866 & 1-94514 & 248.241180 & 42.524670 & 0.0614 & -20.60 & 11.17 & E & 0.51 & 0.00 & --\\
         & 1-210614 & 244.501755 & 41.392189 & 0.0612 & -20.64 & 11.48 & E & 0.49 & 0.01 & 0.40$\pm$0.14\\
1-94784 & 1-211063 & 247.058411 & 40.313835 & 0.0331 & -19.87 & 10.79 & S & 0.33 & 0.09 & 0.20$\pm$0.04\\
         & 1-135502 & 247.764175 & 39.838505 & 0.0305 & -19.51 & 11.13 & S & 0.40 & 0.09 & 0.50$\pm$0.05\smallskip\\
1-44303 & 1-339028 & 116.097923 & 44.527740 & 0.0497 & -20.01 & 11.24 & S & 0.36 & 0.06 & 0.44$\pm$0.08\\
         & 1-379087 & 119.910118 & 51.792362 & 0.0534 & -19.60 & 11.02 & S & 0.38 & 0.10 & 0.72$\pm$0.13\\
1-339094 & 1-274646 & 158.017029 & 43.859268 & 0.0284 & -18.70 & 10.36 & E & 0.53 & 0.02 & 0.35$\pm$0.04\\
         & 1-24099 & 258.027618 & 57.504009 & 0.0282 & -18.67 & 10.34 & E & 0.44 & 0.01 & 0.06$\pm$0.03\smallskip\\
1-137883 & 1-178838 & 312.023621 & 0.068841 & 0.0247 & -17.54 & 10.46 & -- & 0.51 & 0.19 & 0.10$\pm$0.02\\
         & 1-36878 & 42.542126 & -0.867116 & 0.0232 & -18.88 & 10.77 & E & 0.45 & 0.07 & 0.28$\pm$0.04\\
1-48116 & 1-386452 & 136.228333 & 28.384314 & 0.0269 & -19.54 & 10.57 & S & 0.49 & 0.09 & 0.32$\pm$0.04\\
         & 1-24416 & 263.033173 & 56.878746 & 0.0281 & -19.16 & 10.66 & S & 0.37 & 0.03 & 0.22$\pm$0.03\smallskip\\
1-256446 & 1-322671 & 235.797028 & 39.238773 & 0.0637 & -19.77 & 10.82 & E & 0.49 & 0.04 & --\\
         & 1-256465 & 166.752243 & 43.089901 & 0.0575 & -19.70 & 10.79 & E & 0.50 & 0.01 & 0.59$\pm$0.11\\
1-95585 & 1-166947 & 147.335007 & 43.442989 & 0.0720 & -20.79 & 10.81 & S & 0.29 & 0.02 & 0.13$\pm$0.08\\
         & 1-210593 & 244.419754 & 41.899155 & 0.0605 & -19.76 & 10.90 & S & 0.36 & 0.06 & 0.43$\pm$0.14\smallskip\\
1-135641 & 1-635503 & 318.990448 & 9.543076 & 0.0293 & -19.37 & 10.91 & S & 0.22 & 0.10 & 0.15$\pm$0.06\\
         & 1-235398 & 213.149185 & 47.253059 & 0.0281 & -18.91 & 10.99 & S & 0.28 & 0.10 & 0.16$\pm$0.05\\
1-259142 & 1-55572 & 133.121307 & 56.112690 & 0.0454 & -20.11 & 11.03 & S & 0.40 & 0.06 & 0.12$\pm$0.04\\
         & 1-489649 & 171.954834 & 21.386103 & 0.0406 & -19.94 & 10.95 & S & 0.40 & 0.03 & 0.30$\pm$0.08\smallskip\\
1-109056 & 1-73005 & 125.402306 & 45.585476 & 0.0514 & -19.47 & 10.65 & S & 0.31 & 0.05 & 0.20$\pm$0.06\\
         & 1-43009 & 113.553879 & 39.076836 & 0.0510 & -19.41 & 10.43 & S & 0.26 & 0.03 & 0.12$\pm$0.04\\
1-24148 & 1-285031 & 198.701370 & 47.351547 & 0.0303 & -18.47 & 10.72 & S & 0.34 & 0.05 & 0.26$\pm$0.04\\
         & 1-236099 & 225.236221 & 41.566265 & 0.0205 & -17.36 & 9.91 & S & 0.33 & 0.04 & 0.07$\pm$0.01\smallskip\\
1-166919 & 12-129446 & 203.943542 & 26.101791 & 0.0670 & -20.57 & 11.32 & S & 0.34 & 0.03 & 0.28$\pm$0.09\\
         & 1-90849 & 237.582748 & 56.131981 & 0.0661 & -20.39 & 11.16 & E & 0.30 & 0.04 & 0.28$\pm$0.05\\
1-248389 & 1-94554 & 248.914688 & 42.461296 & 0.0318 & -18.96 & 10.57 & S & 0.55 & 0.07 & 0.22$\pm$0.04\\
         & 1-245774 & 214.863297 & 54.100300 & 0.0426 & -20.22 & 10.83 & S & 0.40 & 0.08 & 0.29$\pm$0.07\smallskip\\
1-321739 & 1-247417 & 233.319382 & 45.698528 & 0.0294 & -19.25 & 10.76 & S & 0.28 & 0.09 & 0.16$\pm$0.04\\
         & 1-633994 & 247.419952 & 40.686954 & 0.0305 & -18.27 & 11.04 & S & 0.39 & 0.11 & 0.36$\pm$0.09\\
1-234618 & 1-282144 & 184.592514 & 46.155350 & 0.0492 & -18.92 & 10.31 & S & 0.21 & 0.08 & 0.10$\pm$0.02\\
         & 1-339125 & 117.739944 & 45.989529 & 0.0534 & -18.97 & 11.17 & S & 0.35 & 0.05 & 0.45$\pm$0.23\smallskip\\
1-229010 & 1-210962 & 246.358719 & 39.870697 & 0.0290 & -20.49 & 11.09 & S & 0.47 & 0.07 & 0.35$\pm$0.06\\
         & 1-613211 & 167.861847 & 22.970764 & 0.0323 & -19.87 & 11.32 & E & 0.48 & 0.04 & 0.16$\pm$0.06\\
\hline
\end{tabular}
\label{tablecontrol}
\end{table*}

\begin{table*}
\contcaption{}
\begin{tabular}{ccccccccccccc}
\hline
AGN mangaID & mangaID & RA  & DEC & $z$ & $M_r$ & log $M^\star/M_\odot$ & GZ1$_c$ & $C$ & $A$  & $L(\rm{[OIII]})$\\
 (1)    & (2)      & (3) & (4) & (5) & (6)   & (7)                   & (8)     & (9) & (10) & (11) \\
\hline
1-211311 & 1-25688 & 261.284851 & 58.764687 & 0.0292 & -18.79 & 10.32 & S & 0.29 & 0.06 & 0.10$\pm$0.02\\
         & 1-94422 & 250.453201 & 41.818737 & 0.0316 & -19.15 & 10.55 & S & 0.37 & 0.05 & 0.24$\pm$0.03\smallskip\\
1-373161 & 1-259650 & 196.611053 & 45.289001 & 0.0509 & -21.07 & 11.68 & E & 0.44 & 0.06 & 0.67$\pm$0.20\\
         & 1-289865 & 322.048584 & 0.299885 & 0.0525 & -20.90 & 11.35 & -- & 0.49 & 0.02 & 0.11$\pm$0.09\\
1-210646 & 1-114306 & 323.742737 & 11.296529 & 0.0637 & -20.58 & 10.83 & S & 0.26 & 0.05 & 0.33$\pm$0.16\\
         & 1-487130 & 164.447296 & 21.233431 & 0.0587 & -20.47 & 10.86 & S & 0.26 & 0.11 & 0.27$\pm$0.10\smallskip\\
1-351790 & 1-23731 & 260.746704 & 60.559292 & 0.0205 & -18.20 & 10.19 & E & 0.40 & 0.02 & 0.02$\pm$0.01\\
         & 1-167334 & 151.894836 & 46.093983 & 0.0243 & -18.89 & 10.60 & E & 0.43 & 0.04 & 0.47$\pm$0.05\\
1-163831 & 1-247456 & 232.823196 & 45.416538 & 0.0705 & -20.05 & 10.83 & -- & 0.40 & 0.02 & 0.57$\pm$0.16\\
         & 1-210593 & 244.419754 & 41.899155 & 0.0605 & -19.76 & 10.90 & S & 0.36 & 0.06 & 0.43$\pm$0.14\smallskip\\
1-22301 & 1-251871 & 214.506760 & 41.827644 & 0.1027 & -21.17 & 11.68 & S & 0.26 & 0.05 & 0.24$\pm$0.18\\
         & 1-72914 & 127.580818 & 45.075867 & 0.0970 & -20.88 & 11.31 & S & 0.23 & 0.08 & 0.13$\pm$0.07\\
1-248420 & 1-211063 & 247.058411 & 40.313835 & 0.0331 & -19.87 & 10.79 & S & 0.33 & 0.09 & 0.20$\pm$0.04\\
         & 1-211074 & 247.462692 & 39.766510 & 0.0318 & -19.71 & 10.79 & S & 0.30 & 0.18 & 0.20$\pm$0.04\smallskip\\
1-23979 & 1-320681 & 213.813095 & 47.873344 & 0.0279 & -18.76 & 10.77 & E & 0.48 & 0.01 & 0.09$\pm$0.07\\
         & 1-519738 & 206.514709 & 22.118843 & 0.0277 & -19.49 & 10.73 & E & 0.45 & 0.03 & 0.11$\pm$0.04\\
1-542318 & 1-285052 & 199.061493 & 47.599365 & 0.0573 & -19.77 & 10.85 & S & 0.32 & 0.04 & 0.11$\pm$0.03\\
         & 1-377125 & 112.221359 & 41.307812 & 0.0585 & -19.67 & 10.84 & S & 0.41 & 0.02 & 0.57$\pm$0.14\smallskip\\
1-95092 & 1-210962 & 246.358719 & 39.870697 & 0.0290 & -20.49 & 11.09 & S & 0.47 & 0.07 & 0.35$\pm$0.06\\
         & 1-251279 & 209.251984 & 43.362034 & 0.0329 & -20.11 & 10.97 & E & 0.47 & 0.04 & 0.37$\pm$0.06\\
1-279676 & 1-44789 & 120.890366 & 47.892406 & 0.0586 & -19.33 & 10.92 & -- & 0.31 & 0.13 & 0.32$\pm$0.09\\
         & 1-378401 & 117.904335 & 48.000526 & 0.0612 & -19.65 & 11.02 & E & 0.41 & 0.02 & 0.57$\pm$0.14\smallskip\\
1-201561 & 1-24246 & 264.840790 & 56.567070 & 0.0818 & -19.91 & 10.57 & S & 0.75 & 0.36 & 0.11$\pm$0.06\\
         & 1-285052 & 199.061493 & 47.599365 & 0.0573 & -19.77 & 10.85 & S & 0.32 & 0.04 & 0.11$\pm$0.03\\
1-198182 & 1-256185 & 165.568695 & 44.271709 & 0.0370 & -20.00 & 11.03 & E & 0.50 & 0.06 & 0.25$\pm$0.04\\
         & 1-48053 & 132.595016 & 55.378742 & 0.0308 & -20.24 & 11.49 & E & 0.50 & 0.01 & --\smallskip\\
1-96075 & 1-166947 & 147.335007 & 43.442989 & 0.0720 & -20.79 & 10.81 & S & 0.29 & 0.02 & 0.13$\pm$0.08\\
         & 1-52259 & 59.411037 & -6.274680 & 0.0678 & -20.69 & 11.12 & S & 0.23 & 0.07 & 0.30$\pm$0.09\\
1-519742 & 1-37079 & 42.092335 & 0.986465 & 0.0274 & -17.25 & 9.55 & E & 0.27 & 0.02 & 0.02$\pm$0.01\\
         & 1-276679 & 161.272629 & 44.054291 & 0.0253 & -18.27 & 10.10 & S & 0.24 & 0.03 & 0.05$\pm$0.01\smallskip\\
1-491229 & 1-94554 & 248.914688 & 42.461296 & 0.0318 & -18.96 & 10.57 & S & 0.55 & 0.07 & 0.22$\pm$0.04\\
         & 1-604048 & 50.536137 & -0.836265 & 0.0365 & -20.37 & 10.91 & S & 0.42 & 0.09 & 0.39$\pm$0.08\\
1-604761 & 1-210173 & 241.341766 & 42.488312 & 0.0778 & -20.71 & 11.10 & S & 0.33 & 0.07 & 0.52$\pm$0.13\\
         & 1-71525 & 118.344856 & 36.274380 & 0.0457 & -20.17 & 10.97 & S & 0.27 & 0.10 & 0.19$\pm$0.06\smallskip\\
1-25725 & 1-211079 & 247.438034 & 39.810539 & 0.0304 & -18.97 & 10.54 & E & 0.54 & 0.01 & 0.03$\pm$0.04\\
         & 1-322074 & 228.700729 & 43.665970 & 0.0274 & -18.15 & 10.10 & E & 0.45 & 0.02 & --\\
1-94604 & 1-295095 & 248.348663 & 24.776577 & 0.0410 & -18.40 & 10.14 & E & 0.35 & 0.05 & 0.15$\pm$0.03\\
         & 1-134239 & 241.416443 & 46.846561 & 0.0571 & -19.83 & 10.70 & S & 0.36 & 0.04 & 0.23$\pm$0.06\smallskip\\
1-37036 & 1-210785 & 246.765076 & 39.527386 & 0.0338 & -20.22 & 10.97 & E & 0.47 & 0.01 & --\\
         & 1-25680 & 261.968872 & 60.097275 & 0.0278 & -19.41 & 10.84 & E & 0.52 & 0.02 & 0.34$\pm$0.04\\
1-167688 & 1-235587 & 214.854660 & 45.864250 & 0.0267 & -18.88 & 10.48 & E & 0.44 & 0.01 & 0.08$\pm$0.02\\
         & 1-37062 & 41.846367 & 0.058757 & 0.0248 & -18.30 & 10.40 & E & 0.49 & 0.03 & 0.27$\pm$0.03\smallskip\\
1-279666 & 1-392976 & 156.428894 & 37.497524 & 0.0432 & -17.91 & 10.09 & E & 0.37 & 0.02 & 0.10$\pm$0.03\\
         & 1-47499 & 132.037582 & 54.309921 & 0.0461 & -18.53 & 10.51 & E & 0.27 & 0.04 & 0.15$\pm$0.06\\
1-339163 & 1-136125 & 254.044144 & 34.836521 & 0.0316 & -19.33 & 10.50 & S & 0.25 & 0.09 & 0.08$\pm$0.02\\
         & 1-626830 & 204.683838 & 26.328539 & 0.0282 & -19.23 & 10.67 & S & 0.28 & 0.07 & 0.15$\pm$0.04\smallskip\\
1-258774 & 1-379660 & 119.973717 & 55.374817 & 0.0357 & -19.44 & 10.74 & E & 0.47 & 0.03 & 0.37$\pm$0.07\\
         & 1-48208 & 134.008118 & 57.390965 & 0.0406 & -19.57 & 10.85 & S & 0.50 & 0.01 & 0.12$\pm$0.04\\
1-198153 & 1-211063 & 247.058411 & 40.313835 & 0.0331 & -19.87 & 10.79 & S & 0.33 & 0.09 & 0.20$\pm$0.04\\
         & 1-135810 & 250.123138 & 39.235115 & 0.0297 & -19.38 & 10.59 & S & 0.24 & 0.09 & 0.08$\pm$0.02\smallskip\\
1-91016 & 1-338828 & 115.641609 & 44.215858 & 0.0418 & -18.10 & 10.42 & S & 0.28 & 0.03 & 0.43$\pm$0.05\\
         & 1-386695 & 137.983505 & 27.899269 & 0.0474 & -19.33 & 10.48 & S & 0.27 & 0.09 & 0.81$\pm$0.09\\
1-279073 & 1-211100 & 247.830322 & 39.744129 & 0.0309 & -19.15 & 10.62 & E & 0.56 & 0.02 & --\\
         & 1-210784 & 247.097122 & 39.570305 & 0.0292 & -19.61 & 10.86 & E & 0.48 & 0.00 & 0.15$\pm$0.05\smallskip\\
1-135044 & 1-218280 & 124.003311 & 27.075895 & 0.0255 & -19.57 & 10.81 & S & 0.27 & 0.08 & 0.12$\pm$0.03\\
         & 1-211063 & 247.058411 & 40.313835 & 0.0331 & -19.87 & 10.79 & S & 0.33 & 0.09 & 0.20$\pm$0.04\\
1-148068 & 1-166947 & 147.335007 & 43.442989 & 0.0720 & -20.79 & 10.81 & S & 0.29 & 0.02 & 0.13$\pm$0.08\\
         & 1-55572 & 133.121307 & 56.112690 & 0.0454 & -20.11 & 11.03 & S & 0.40 & 0.06 & 0.12$\pm$0.04\\
\hline
\end{tabular}
\end{table*}

\begin{table*}
\contcaption{}
\begin{tabular}{ccccccccccccc}
\hline
AGN mangaID & mangaID & RA  & DEC & $z$ & $M_r$ & log $M^\star/M_\odot$ & GZ1$_c$ & $C$ & $A$  & $L(\rm{[OIII]})$\\
 (1)    & (2)      & (3) & (4) & (5) & (6)   & (7)                   & (8)     & (9) & (10) & (11) \\
\hline
1-277552 & 1-264513 & 236.941513 & 28.641697 & 0.0333 & -20.92 & 11.28 & S & 0.25 & 0.18 & 0.33$\pm$0.05\\
         & 1-136125 & 254.044144 & 34.836521 & 0.0316 & -19.33 & 10.50 & S & 0.25 & 0.09 & 0.08$\pm$0.02\\
1-217050 & 1-135372 & 250.116714 & 39.320118 & 0.0301 & -20.29 & 11.08 & E & 0.49 & 0.02 & 0.01$\pm$0.23\\
         & 1-274663 & 157.660522 & 44.012722 & 0.0280 & -19.88 & 11.00 & E & 0.50 & 0.01 & 0.08$\pm$0.02\smallskip\\
1-25554 & 1-135625 & 248.507462 & 41.347946 & 0.0284 & -19.06 & 10.56 & S & 0.43 & 0.05 & 0.56$\pm$0.04\\
         & 1-216958 & 136.200287 & 40.591721 & 0.0270 & -18.95 & 10.41 & S & 0.51 & 0.03 & 0.23$\pm$0.02\\
1-135285 & 1-633990 & 247.304123 & 41.150871 & 0.0296 & -19.06 & 10.46 & S & 0.34 & 0.03 & 0.25$\pm$0.03\\
         & 1-25688 & 261.284851 & 58.764687 & 0.0292 & -18.79 & 10.32 & S & 0.29 & 0.06 & 0.10$\pm$0.02\\
\hline
\end{tabular}
\end{table*}

% Don't change these lines
\bsp	% typesetting comment
\label{lastpage}

\begin{thebibliography}{99}
\bibitem[\protect\citeauthoryear{Abolfathi et al.}{2018}]{dr14} Abolfathi, B. et al. ApJS, 2018, 235, 42

\bibitem[\protect\citeauthoryear{Antonucci}{1993}]{antonucci93} 
Antonucci, R., 1993, ARAA, 31, 473

\bibitem[\protect\citeauthoryear{Bower et al.}{2006}]{bower06} 
Bower, R. G. et al. 2006, MNRAS, 370, 645

\bibitem[\protect\citeauthoryear{Barbosa et al.}{2014}]{barbosa14}
	Barbosa, F. K. B., Storchi-Bergmann, T., McGregor, P., Vale, T. B.,Riffel, R. A., 2014, MNRAS, 445, 2353
	
\bibitem[\protect\citeauthoryear{Bundy et al.}{2015}]{bundy15}
Bundy, K., et al., 2015, ApJ 798,  7

\bibitem[\protect\citeauthoryear{Bruzual \& Charlot}{2003}]{bc03}  
Bruzual G., Charlot S., 2003,   
% \textsl{Stellar population synthesis at the resolution of 2003},  
MNRAS,  344, 1000 


\bibitem[\protect\citeauthoryear{Cappellari \& Emsellem}{2004}]{cappellari04} 
Cappellari, M., Emsellem, E., 2004, PASP, 116, 138

\bibitem[\protect\citeauthoryear{Cappellari et al.}{2007}]{cappellari07}
Cappellari, M. et al., 2007, MNRAS, 379, 418

\bibitem[\protect\citeauthoryear{Cappellari}{2017}]{cappellari17}
Cappellari, M., 2017, MNRAS, 466, 798

\bibitem[\protect\citeauthoryear{Cheung et al.}{2016}]{cheung16} Cheung, E., et al., 2016, Nature, 533, 504.

\bibitem[\protect\citeauthoryear{Conroy, Gunn \& White.}{2009}]{Conroy+09}
Conroy, C., Gunn, J. E., White, M., 2009, ApJ, 699, 486

\bibitem[\protect\citeauthoryear{Di Matteo et al.}{2005}]{diMatteo05}
Di Matteo, T., Springel, V. \& Hernquist, L. 2005, Nature, 433, 604

\bibitem[\protect\citeauthoryear{Drory et al.}{2015}]{drory15}  
Drory, N., et al. 2015, AJ, 149, 77  

\bibitem[\protect\citeauthoryear{Ferrarese \& Merritt}{2000}]{ferrarese00} 
Ferrarese, L. \& Merritt, D. A., 2000, ApJL, 539, L9
\bibitem[\protect\citeauthoryear{Fischer et al.}{2013}]{fischer13} 
Fischer, T. C., Crenshaw, D. M., Kraemer, S. B., Schmitt, H. R.	 2013, ApJS, 209, 1
\bibitem[\protect\citeauthoryear{Gebhardt et al.}{2000}]{gebhardt00} 
Gebhardt, K. et al. 2000, ApJL, 539, L13

\bibitem[\protect\citeauthoryear{Gonzalez \& Woods}{2002}]{gonzalez02} 
Gonzalez R.C., Woods R.E., 2002, Digital Image Processing, 2nd edition, Prentice-Hall, Englewood Cliffs, NJ

\bibitem[\protect\citeauthoryear{Hopkins et al.}{2005}]{hopkins05} 
Hopkins, P. et al. 2005, ApJ, 630, 705

\bibitem[\protect\citeauthoryear{Krajnovi\'c et al.}{2005}]{krajnovic05}
Krajnovi\'c, D., Cappellari, M., Emsellem, E., McDermid, R. M., de Zeeuw, P. T., 2005, 357, 1113
\bibitem[\protect\citeauthoryear{Krajnovi\'c et al.}{2006}]{krajnovic06} 
Krajnovi\'c, D., Cappellari, M., de Zeeuw, P. T., Copin, Y., 2006, MNRAS, 366, 787.
\bibitem[\protect\citeauthoryear{Krajnovi\'c et al.}{2011}]{krajnovic11}
Krajnovi\'c, D. et al., 2011, MNRAS, 414, 2993


\bibitem[\protect\citeauthoryear{Law et al.}{2015}]{law15}
Law, D. R., et al., 2015, AJ, 150, 19
\bibitem[\protect\citeauthoryear{Law et al.}{2016}]{law16}
Law, D. R., et al., 2015, AJ, 152, 83

\bibitem[\protect\citeauthoryear{Lena et al.}{2015}]{lena15}
Lena, D. et al., 2015, ApJ, 806, 84

%\bibitem[\protect\citeauthoryear{Menezes, Steiner \& Ricci}{2014}]{menezes14}
%Menezes, R. B., Steiner, J. E., Ricci, T. V., 2014, MNRAS, 438, 2597.	

%\bibitem[\protect\citeauthoryear{Menezes, Steiner \& Ricci}{2015}]{menezes15} 
%Menezes, R. B.,  da Silva, P., Ricci, T. V., Steiner, J. E., May, D., Borges, B. W., 2015, MNRAS, 450, 369.	

\bibitem[\protect\citeauthoryear{Mallmann et al.}{2018}]{nicolas18} 
Mallmann, N. D. et al. 2018, MNRAS, submitted


\bibitem[\protect\citeauthoryear{Nemmen et al.}{2007}]{nemmen07}
Nemmen R., Bower, R., Babul, A. \& Storchi-Bergmann, T. 2007, MNRAS, 377, 1652

\bibitem[\protect\citeauthoryear{Oh et al.}{2011}]{oh11} 
	Oh, K., Sarzi, M., Schawinski, K., Yi, S. K., 2011, ApJS, 195, 130

\bibitem[\protect\citeauthoryear{Osterbrock \& Ferland}{2006}]{osterbrock06} Osterbrock, D. E., \& Ferland, G. J. 2006, Astrophysics of Gaseous Nebulae and Active Galactic Nuclei, 2nd ed., University Science Books, Sausalito, California

\bibitem[\protect\citeauthoryear{Penny et al.}{2018}]{penny18}
Penny, S. J. et al, 2018, MNRAS, submitted

\bibitem[\protect\citeauthoryear{Rembold et al.}{2017}]{rembold17}
Rembold, S. B. et al, 2017, MNRAS, 472, 4382

\bibitem[\protect\citeauthoryear{Riffel et al.}{2014}]{n5929let}
	Riffel, R. A., Storchi-Bergmann, T., Riffel, R., 2014, ApJ, 780, 24

\bibitem[\protect\citeauthoryear{Sarzi et al.}{2006}]{sarzi06}
Sarzi, M., et al, 2006, MNRAS, 366, 1151

\bibitem[\protect\citeauthoryear{Scannapieco et al.}{2005}]{scannapieco05} 
Scannapieco, E. et al. 2005, ApJ, 635, L13

\bibitem[\protect\citeauthoryear{Schmitt et al.}{2003}]{schmitt03} 
Schmitt, H. R., Donley, J. L., Antonucci, R. R. J., Hutchings, J. B., Kinney, A. L., 2003, ApJS, 148, 327

\bibitem[\protect\citeauthoryear{Schnorr-M\"uller et al.}{2014}]{allan14} 
	Schnorr-M\"uller, A., Storchi-Bergmann, T., Nagar, N. M., Robinson, A., Lena, D., Riffel, R. A., Couto, G. S., 2014, MNRAS, 437, 1708
    
\bibitem[\protect\citeauthoryear{Springel et al.}{2005}]{springel05} 
Springel, V., Di Matteo, T. \& Hernquist, L. 2005, ApJ, 620, L79

\bibitem[\protect\citeauthoryear{Tremaine et al.}{2002}]{tremaine02}
Tremaine, S. et al. 2002, ApJ, 574, 740

\bibitem[\protect\citeauthoryear{Urry \& Padovani}{1995}]{urry95} 
Urry, C. M., Padovani, P., 1995, PASP, 107, 803

\bibitem[\protect\citeauthoryear{Veilleux et al.}{2013}]{veilleux13} 
Veilleux S., et al., 2013, ApJ, 776, 27

\bibitem[\protect\citeauthoryear{Wake  et al.}{2017}]{wake17} 
Wake, D. A. et al., 2017, AJ, 154, 86

\bibitem[\protect\citeauthoryear{Woo et al.}{2017}]{woo17}
Woo, J., et al., 2017, ApJ, 839, 120

\bibitem[\protect\citeauthoryear{Wylezalek et al.}{2018}]{wylezalek18}
Wylezalek, D. et al, 2018, MNRAS, 474, 1499

\bibitem[\protect\citeauthoryear{Wylezalek et al.}{2017}]{wylezalek17}
Wylezalek, D. et al, 2017, MNRAS, 467, 2612

\bibitem[\protect\citeauthoryear{Wylezalek et al.}{2016}]{wylezalek16}
Wylezalek, D. et al, 2016, MNRAS, 461, 3724

\bibitem[\protect\citeauthoryear{Yan et al.}{2016a}]{yan16}
Yan, R., et al., 2016a, AJ, 151, 8

\bibitem[\protect\citeauthoryear{Yan et al.}{2016b}]{yan16b}
Yan, R., et al., 2016b, AJ, 152, 197


\bibitem[\protect\citeauthoryear{Zakamska et al.}{2016}]{zakamska16}
Zakamska, N. L. et al, 2016, MNRAS, 459, 3144

\bibitem[\protect\citeauthoryear{Zakamska \& Greene}{2014}]{zakamska14}
Zakamska N. L., Greene J. E., 2014, MNRAS, 442, 784

\bibitem[\protect\citeauthoryear{Zubovas \& King}{2012b}]{zubovas12b}
Zubovas K., King A., 2012b, ApJ, 745, L34
\end{thebibliography}
\end{document}